\documentclass[aps,prd,superscriptaddress,showkeys,showpacs,twocolumn,a4paper]{revtex4-2}
\usepackage{ulem,amssymb,MnSymbol}
\usepackage{cancel}
\usepackage{subcaption}
\usepackage{graphicx}
\usepackage{epsfig}
\usepackage{mathrsfs}
\usepackage{amsmath,amsthm}
\usepackage[dvipsnames]{xcolor}
\usepackage[bottom]{footmisc}
\definecolor{darkblue}{RGB}{0,0,196}
\definecolor{darkgreen}{RGB}{0,120,0}
\usepackage[colorlinks=true,linktocpage=true,linkcolor=darkblue,citecolor=red,urlcolor=darkblue]{hyperref}
\usepackage{orcidlink}
\usepackage{ragged2e}

\newcommand{\be}{\begin{equation}}
\newcommand{\ee}{\end{equation}}
\newcommand{\ba}{\begin{eqnarray}}
\newcommand{\ea}{\end{eqnarray}}

\begin{document}
\title{Rotating magnetized pion gas of finite transverse size: condensation constraints and transport properties}
\author{Ankit Kumar\,\orcidlink{0009-0004-2988-9269}}
\email{kumar.ankit@iitgn.ac.in}
\affiliation{Indian Institute of Technology Gandhinagar, Gandhinagar-382355, Gujarat, India}

\author{Diwakar Gaur\,\orcidlink{0009-0006-5358-5753}}
\email{diwakar.gaur@iitgn.ac.in}
\affiliation{Indian Institute of Technology Gandhinagar, Gandhinagar-382355, Gujarat, India}

\author{Vinod Chandra\, \orcidlink{0000-0002-2075-8041}}
\email{vchandra@iitgn.ac.in}
\affiliation{Indian Institute of Technology Gandhinagar, Gandhinagar-382355, Gujarat, India}

\begin{abstract}
This work investigates the electric, thermal, and thermoelectric responses of a rotating pion gas of finite transverse radius in the presence of a background magnetic field, with the rotation axis aligned with the magnetic field. We explicitly calculate the parameter limits for $\pi^+$ condensation and restrict our working regime safely outside these boundaries, ensuring well-behaved transport coefficients. Notably, the system exhibits a condensation asymmetry, with $\pi^-$ remaining uncondensed at the parameters that induce $\pi^+$ condensation. Using the Boltzmann Transport Equation under the Relaxation Time Approximation, we calculate the longitudinal electrical conductivity, thermal conductivity, and the Seebeck coefficient. Our results reveal a competing interplay between the magnetic field and rotation, highlighting the substantial impact of rotation on the medium's transport properties: while the magnetic field suppresses the transport coefficients in a static medium, rotation, acting as an effective chemical potential, introduces an energy shift that favors their increase. Beyond an angular velocity, this rotational enhancement overpowers the magnetic suppression, leading to an increase in the transport coefficients with increasing magnetic field. Finally, we analyze the relative significance of charge and heat transport through the Lorenz number, providing further insight into the transport characteristics of the rotating magnetized pion medium.
\end{abstract}

\maketitle
  \section{Introduction}\label{secI}

The study of strongly interacting matter under extreme conditions has been a focal point of high-energy physics, particularly in relativistic Heavy-Ion Collisions (HICs). These collisions create a hot, dense medium of quarks and gluons known as Quark-Gluon Plasma (QGP), which subsequently expands, cools, and evolves into a confined hadronic phase.  In this late stage of the collision evolution, pions emerge as the lightest and the most abundant hadronic species \cite{PhysRevC.88.044910,PhysRevC.96.044904,ANDRONIC2006167}. Consequently, understanding the transport properties of the pion gas is of fundamental importance. The transport coefficients are required to study the hydrodynamical evolution and properties of the pionic medium and allow for bridging theoretical models with experiments in the post-hadronization regime. Transport coefficients of hot pion gas, such as conductivities and viscosities, have been extensively studied through various frameworks, including kinetic theory \cite{PRAKASH1993321,PhysRevD.89.054013,PhysRevD.93.096012}, chiral perturbation theory \cite{fernandez2009transport}, holographic methods \cite{dobado2007ratio}, and the Kubo formalism \cite{lang2012shear}.

In non-central HICs, the geometric overlap of the colliding nuclei is highly asymmetric, leading to two very important physical phenomena. First, the spectator nucleons that do not participate directly in the collision but move at velocities very close to the speed of light, generate extraordinarily strong, transient background magnetic fields of order $eB\sim m_{\pi}^2$ perpendicular to the reaction plane \cite{KHARZEEV2008227,Kharzeev_2016,Skokov:2009qp,inghirami2020magnetic}. Such extreme fields heavily influence the transport dynamics of the medium, as the transverse motion of the charged particles becomes quantized into discrete Landau Levels (LL) \cite{KHARZEEV2008227,zahed,PhysRevLett.120.162301,kurian2019transport,PhysRevC.107.034903}. Although these fields are transient, theoretical models suggest that finite electrical conductivity of the QGP medium created in these collisions can extend the lifetime of the fields to even beyond the hadronization phase \cite{PhysRevC.107.034901,PhysRevC.82.034904,https://doi.org/10.1155/2013/490495,STEWART2021122308}. Second, these collisions deposit a massive amount of angular momentum of the order of $10^3-10^7\,\,\hbar$ into the interacting region \cite{star2017global,PhysRevC.96.054908,Becattini_2008,PhysRevC.93.064907,Jiang_2016,PhysRevC.85.054901}. This initial angular momentum is converted into significant local vorticity within the QGP and subsequent hadronic phases, which can cause polarization in the particles produced in the collision \cite{PhysRevC.97.041901,BECATTINI2021136519}. Measurements of the global polarization of $\Lambda$ hyperons by the STAR collaboration \cite{star2017global} provide strong experimental evidence for this retained angular momentum. Since pions are spin-$0$ particles, they couple to the global rotation purely through orbital angular momentum. 

The simultaneous presence of a magnetic field and rotation strongly affects the energy spectrum of quantum fields. For charged pions modeled as complex scalar fields in a rotating frame, the rotation couples to their angular momentum, effectively acting as a chemical potential \cite{zahed}. Theoretical investigations into the combined effects of parallel rotation and magnetic fields have demonstrated that rotation can linearly shift and lift the degeneracy of the LLs; this rotation shift can lead to the condensation of charged pions first discussed in \cite{zahed} and further explored in \cite{m97n-qxck,PhysRevD.100.094015,chen2024charged,Zhang_2020,PhysRevD.109.056024}.
While the individual effects of either magnetic fields or rotation on the properties of hadronic matter have been independently examined in some detail \cite{padhan,PhysRevD.102.014030,PhysRevC.107.034903,pradhan2024thermodynamics,mukherjee2024conserved,kalikotay,kalikotay2024electrical,h9kt-bj75}, their combined effect on the transport of a hot pion gas remains largely unexplored. Addressing this theoretical gap forms the primary motivation for the present study.

In this work, we investigate the electric, thermal, and thermoelectric responses of a rotating pion gas in a constant background magnetic field, with the axis of rotation aligned with the magnetic field. While the previous studies have investigated the pionic medium using infinite-volume energy spectra, we focus on how imposing a finite transverse boundary condition modifies the system's properties. We model charged pions as complex scalars and neutral pions as real scalars within a rotating frame. By evaluating the pion condensation thresholds discussed in Ref.~\cite{zahed} under the constraint of a finite system radius, we explicitly restrict our chosen parameter space to a regime where such condensation is avoided. Consequently, our calculations of the transport coefficients remain strictly valid within this defined domain. Using the Boltzmann Transport Equation (BTE) under the Relaxation Time Approximation (RTA), we compute the longitudinal electrical conductivity, thermal conductivity, Seebeck coefficient, and Lorenz number, and study the effect of the magnetic field and rotation on these transport coefficients. We restrict ourselves only to the longitudinal components of the transport coefficients due to the Landau quantization in the transverse plane. In our considered system, the total electrical charge is strictly conserved; therefore, the system is described using a finite charge chemical potential $\mu$. Our analysis reveals a competition between magnetic-field suppression and rotational enhancement.
In a static medium, the magnetic field suppresses transport; however, rotation acts as an effective chemical potential that introduces a favorable energy shift. Once the angular velocity is sufficiently large, this rotational enhancement dominates the magnetic suppression, resulting in transport coefficients that increase with the magnetic field. The ratio of the Lorenz number (related to the ratio of thermal to electrical conductivities) in the rotating, magnetized medium to its zero-field, non-rotating counterpart highlights the non-trivial interplay between magnetic field and rotation in governing the balance between thermal and electrical transport.

The paper is organized as follows. In Sec.~\ref{spectra}, we present the energy spectra for charged and neutral pions confined by a finite transverse boundary in a rotating frame and introduce the corresponding equilibrium distribution functions. In Sec.~\ref {PC}, we analyze pion condensation using the charged-pion spectrum within the causal domain. Sec.~\ref{TC} details the derivation of the transport coefficients using the BTE within RTA. Finally, in Sec.~\ref {RD}, we present our results, followed by a conclusion and possible future directions in Sec.~\ref {outlook}.

\section{Spectra and equilibrium distribution functions }\label{spectra}
In this section, we establish the energy spectra and equilibrium distribution functions for both charged and neutral pions in a rotating frame with a background magnetic field. We briefly review the infinite-volume charged pion spectrum discussed in Ref.~\cite{zahed}. Building on this foundation, we impose physically realistic finite-radius boundary conditions to derive the modified charged pion spectrum and subsequently extend our formulation to include neutral pions. We consider the metric for a rotating frame in 1+3 dimensions with $(+, - , - , -)$ signature, such that the axis of rotation is along $\hat{\mathbf{z}}$:
\begin{align}\label{metric}
    ds^2 &=g_{\mu\nu}dx^{\mu}dx^{\nu}\\\nonumber
    &= (1-\Omega^2 \rho^2)dt^2 + 2 y \Omega dx dt - 2 x \Omega dy dt - dr^2,\\\nonumber
    g_{\mu\nu}&=\begin{pmatrix}
        1-\Omega^2\rho^2 & \Omega y & -\Omega x & 0\\
        \Omega y & -1 & 0 & 0\\
        -\Omega x & 0 & -1 & 0\\
        0 & 0 & 0 & -1
    \end{pmatrix}
\end{align}
where, $\rho^2 = x^2 + y^2$, and $\Omega$ is the angular velocity of rotation. Causality demands that $v=\Omega R<1$, with $R$ being the radius of the system. A constant magnetic field, $\mathbf{B} = B \hat{\mathbf{z}}$, is  described by the circular vector potential, which is given in a rotating frame as:
\begin{equation}
    A_\mu = B (-\frac{\Omega \rho^2}{2},\frac{y}{2},-\frac{x}{2},0)
\end{equation}
 \\
In the absence of rotation, the spectrum of charged pions in an infinite volume subjected to a uniform magnetic field is characterized by highly degenerate Landau Levels, labeled by $n$:
\begin{equation}\label{infinitespectra}
    E_{n}^j(p_{z})=\sqrt{|eB|(2n+1)+p_z^2+m_{\pi}^2}
\end{equation}
where $p_z$ is the pion momentum along $z$-direction and $j = \pm1$ for $\pi^{\pm}$. The degeneracy is given by $ N = \frac{|eB|A}{2 \pi}$, $A = \pi R^2$ being the area transverse to $\mathbf{B}$. 
A rotation $\Omega$ parallel to $\mathbf{B}$ shifts the spectrum linearly, lifting the degeneracy of each Landau level:
\begin{equation}
    E_n^j(p_z) \rightarrow E_{n,l}^j (p_z) \equiv E_n^j(p_z) - \Omega L_z ={E_n^j(p_z) - j\Omega l}
\end{equation}
 $jl$ is the angular momentum $L_z$ for $\pi^{\pm}$, with the restriction $-n\le l\le N-n$. 
 
 However, for a system with a finite radius, the mode function vanishes at the boundary. Under this constraint, the equation of motion for a charged scalar field $\Pi$ in a rotating frame is given by
 \begin{align}
     -(\partial_t+y\Omega\partial_x-x\Omega\partial_y)^2\Pi-D^{\dagger}_iD_i\Pi-m^2\Pi=0,
 \end{align}
 where $D_{\mu}=\partial_{\mu}+ieA_{\mu}$, can be solved with cylindrical mode functions of the form
\begin{equation}
\begin{split}
     \psi_{l,p_z,k}(t,\rho,\phi,z) = e^{-iEt}e^{i p_z z}e^{i l \phi} \rho^{|l|} e^{-eB \rho^2/4} 
     \\
    \times\,\, {_1F_1}\Big(-a,|l|+1,\frac{ e B \rho^2}{2}\Big),
\end{split}
\end{equation}
where $_1F_1$ is the Kummer confluent hypergeometric function with parameter $a$ given by
\begin{align}\label{al}
    -a(l)=\frac{1}{2}(|l|-l+1)-\frac{1}{2eB}((E+\Omega l)^2-p_z^2-m^2),
\end{align}
and $l\in \mathbb{Z}$. The boundary condition requires
\begin{equation}
    _1F_1\Big(-a(l),|l|+1,\frac{ e B R^2}{2}\Big) = 0,
\end{equation}
which fixes $a$. With this, and Eq. \eqref{al}, one can write the spectra for the charged pions as
\begin{align}\label{E}
    E_{l,k,j}&(p_z) =p_{0j}\\\nonumber
    &= \sqrt{m^2+p_z^2 + e B(2 a_k(l)+|l|-l+1) }-j\Omega l 
\end{align}
where $a_k(l)$ is the $k^{\text{th}}$ root of the $_1F_1$ function. 

Neutral pions can be modeled as real scalar fields. Scalar fields in rotating coordinates have been studied previously in various contexts \cite{letaw, duffy, ambrus, Bordag2025}.
The Klein-Gordon equation in rotating coordinates is
\begin{align}\label{KGn}
    -(\partial_t+y\Omega\partial_x-x\Omega\partial_y)^2\psi+\nabla^2\psi-m^2\psi=0.
\end{align}
Mode functions that satisfy this equation are
\begin{align}
    \psi(t,\rho,\phi,z)=e^{-iEt}e^{ip_zz}e^{il\phi}J_l(p_{\perp}\rho),
\end{align}
with the restriction that  
\begin{align}\label{spectrum0}
    p_{\perp}^2=(E+\Omega l)^2-p_z^2-m^2.
\end{align}
Here $J_l$ are the Bessel functions, and $l\in\mathbb{Z}$. In a system of finite radius, mode functions vanish at the boundary, i.e., $J_l(p_{\perp}R)=0$, which implies
\begin{align}
    p_\perp = \frac{\alpha_k(l)}{R}
\end{align}
where $\alpha_k(l)$ is the $k^{\text{th}}$ root of the Bessel function $J_l$. Using this and Eq. \eqref{spectrum0}, we obtain the spectrum of neutral pions as
\begin{align}
    E_{l,k}(p_z) =p_{0u}= \sqrt{m^2+p_z^2+\Big(\frac{\alpha_k(l)}{R}\Big)^2} - \Omega l.
\end{align}
 
In Local Rest Frame (LRF), $u^{\mu}=(\frac{1}{\sqrt{1-\Omega^2\rho^2}},0,0,0)$. The equilibrium distribution function for the charged pions in the LRF can then be written as
\begin{align}\label{f0j}
        f^0_j &= \frac{1}{e^{\beta(p_{\alpha}u^{\alpha}-j\mu)}-1}=\frac{1}{e^{\beta\Big(\frac{p_{0j}}{\sqrt{1-\Omega^2\rho^2}}-j \mu\Big)}-1}
\end{align}
where $\mu$ is the charge chemical potential. For neutral pions:
\begin{equation}\label{f0u}
    f^0_u = \frac{1}{e^{\Big(\frac{\beta p_{0u}}{\sqrt{1-\Omega^2\rho^2}}\Big)}-1}.
\end{equation}
The $u$ in the distribution function's subscript stands for uncharged. 

\section{Condensation constraints}\label{PC}
Rotation can linearly shift the Landau levels, triggering pion condensation at a specific critical angular velocity \cite{zahed}. In this section, we investigate how the constraint of a finite system radius modifies this condensation threshold in the presence of a finite charge chemical potential $\mu$. The lowest energy state for a given $l$ is
\begin{align}
    E_l=\sqrt{m^2+eB(2 a_1(l)+|l|-l+1)}-j \Omega l
\end{align}
Let $E_{l^*}$ be the minimum energy, where $l^*$ denotes the lowest energy state, then the condition for condensation is:
\begin{align}
    E_{l^*} -j \mu \sqrt{1-\Omega^2\rho^2}= 0
\end{align}
The magnitude of the chemical-potential term is largest at $\rho = 0$. For $\pi^+$ ($j = +1$), this term lowers the left-hand side and thus favors condensation; its effect is strongest at the center and weakens with increasing $\rho$. Consequently, condensation sets in most readily at $\rho = 0$, where the chemical potential required to trigger it is smallest: for any $\rho \neq 0$, a larger $\mu$ would be needed. To identify the limiting parameters for condensation, it is therefore sufficient to evaluate the condition at $\rho = 0$. For $\pi^-$ ($j = -1$), the same term enters with a positive sign and thus only raises the threshold, never favoring condensation. Therefore, later in this section, to determine whether condensation occurs within our parameter range, we drop this term entirely and study $E_{l^*}$ alone.
Setting $\rho = 0$, the condensation condition for $\pi^+$ becomes,
\begin{align}
    \sqrt{m^2+eB(2a_1(l^*)+1)}-\Omega l^*=\mu_c
\end{align}

\begin{figure} [h] 
        \centering
        \includegraphics[width=0.45\textwidth]{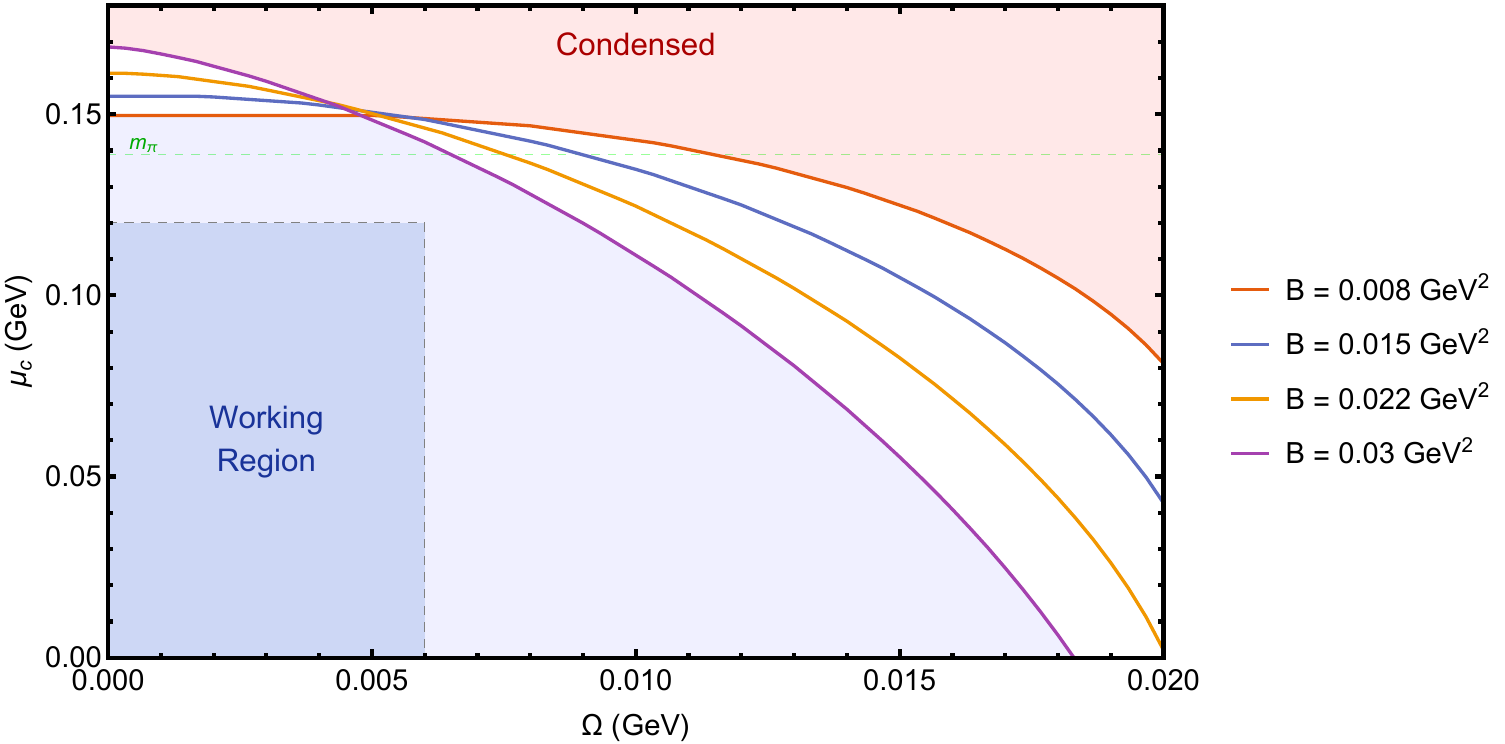}
        \caption{\justifying \small Threshold chemical potential $\mu_c$ for $\pi^+$ with rotation $\Omega$ for different values of magnetic field at $R = 50\,\,  \text{GeV}^{-1}\approx10\,\,\text{fm}$ \cite{zahed}. The blue-shaded region shows the parameter range over which we evaluate the transport coefficients. The green dotted line represents the mass scale of the pions.}
        \label{omegac}      
\end{figure}     
\begin{figure*}[ht]
    \centering
    \begin{subfigure}{0.4\textwidth}
        \centering
        \includegraphics[width=\textwidth]{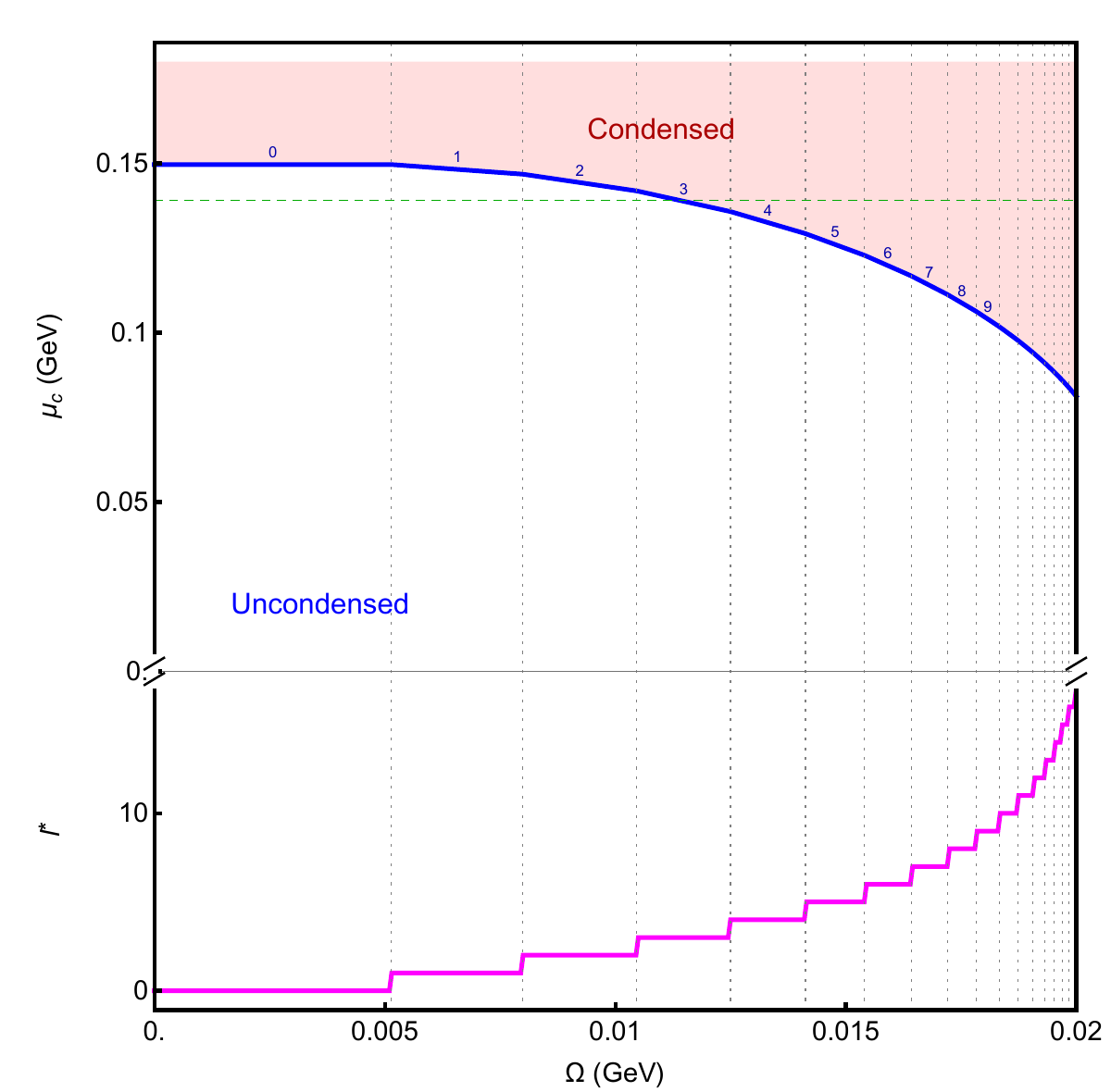} 
        \caption{\small $B=0.008\,\,\text{GeV}^2$}
        \label{dotted1}
    \end{subfigure}
    \begin{subfigure}{0.4\textwidth}
        \centering
        \includegraphics[width=\textwidth]{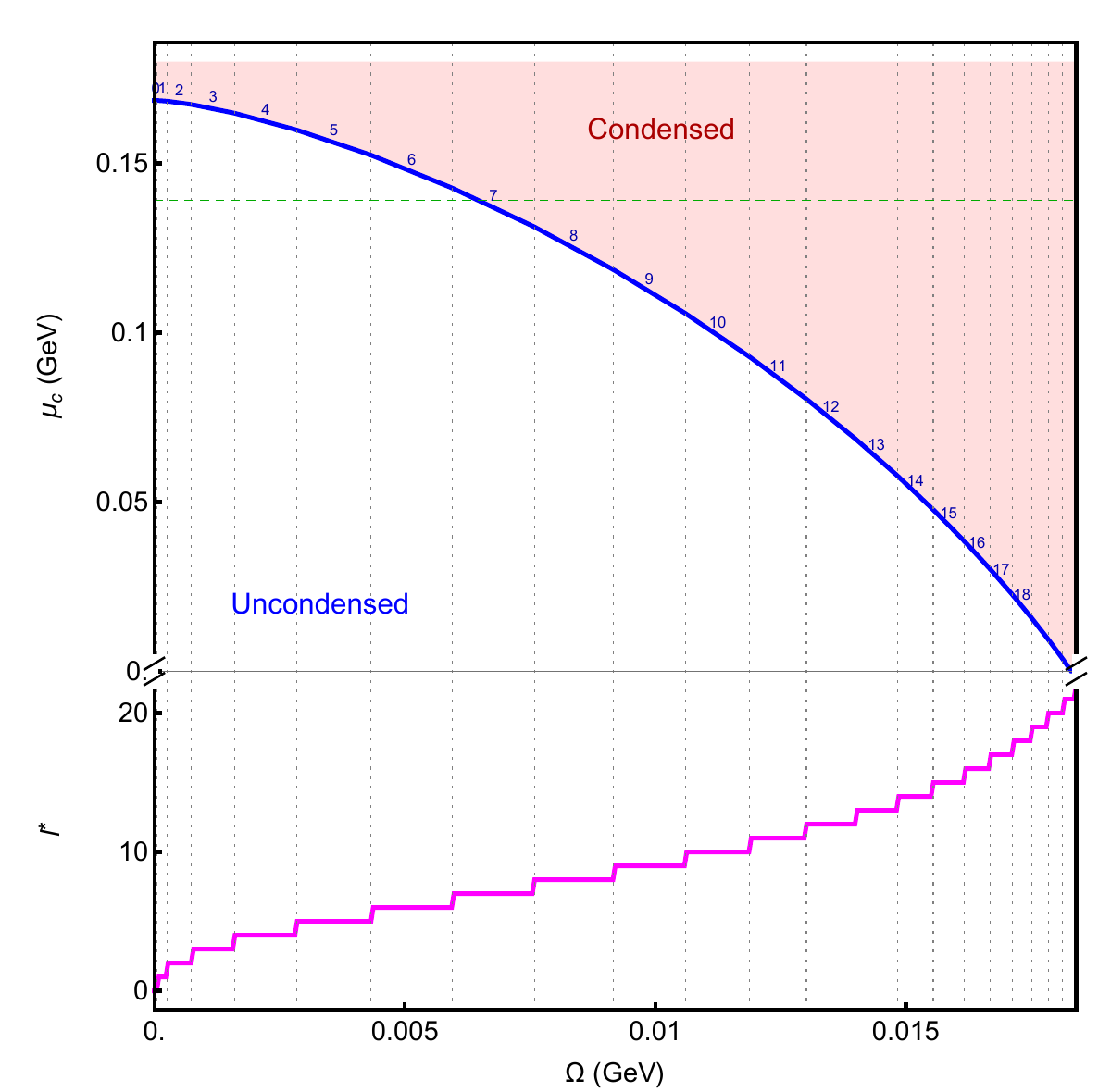} 
        \caption{\small $B=0.03\,\,\text{GeV}^2$}
        \label{dotted2}
    \end{subfigure}
\caption{\justifying \small $\mu_c$ (upper panel) and $l^*$ (lower panel) for $\pi^+$ with $\Omega$ at R = 50  GeV$^{-1}$. This figure illustrates how the lowest energy state of $\pi^+$ varies with $\Omega$ for a given value of $B$. The numbers on the blue line indicate $l^*$ in that region. }
\label{dotted}
\end{figure*}
\begin{figure}[b]
    \centering
    \includegraphics[width=0.93\linewidth]{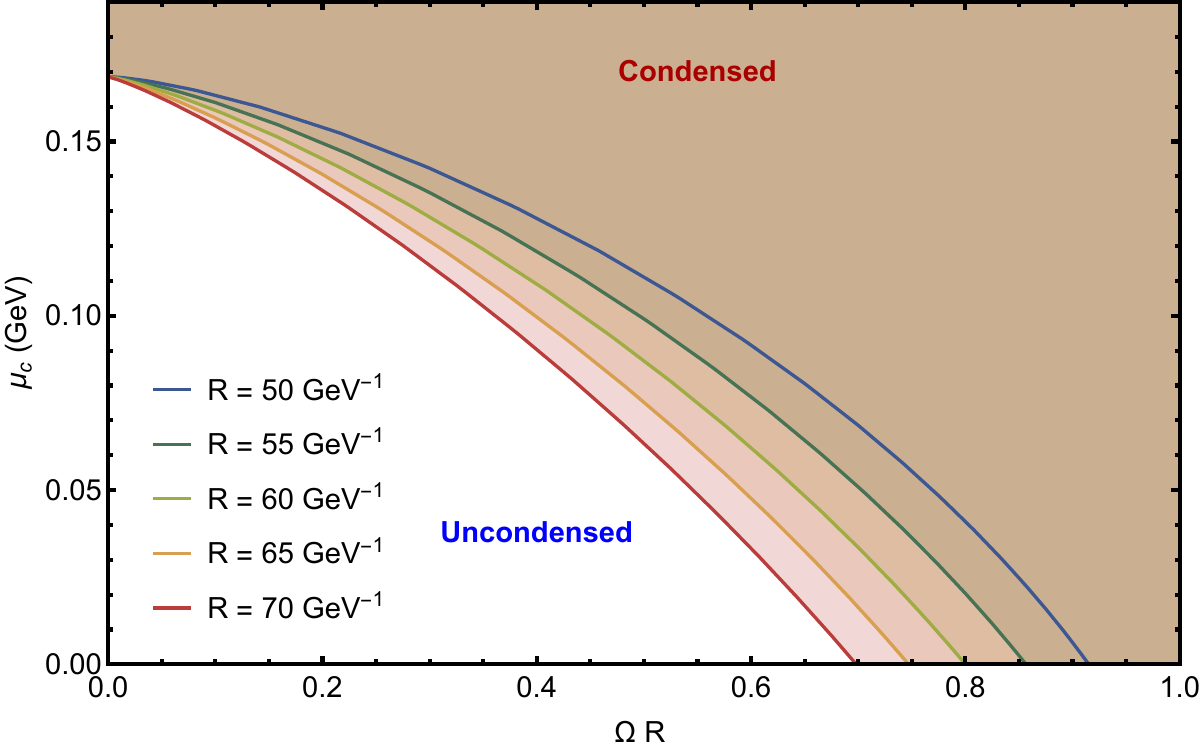}
    \caption{\justifying \small $\mu_c$ for $\pi^+$ with $\Omega R$ for different values of the system radius $R$ at $B=0.03\,\,\text{GeV}^2$.}
    \label{rcurve}
\end{figure}
\begin{figure}[b]
    \centering
    \includegraphics[width=0.9\linewidth]{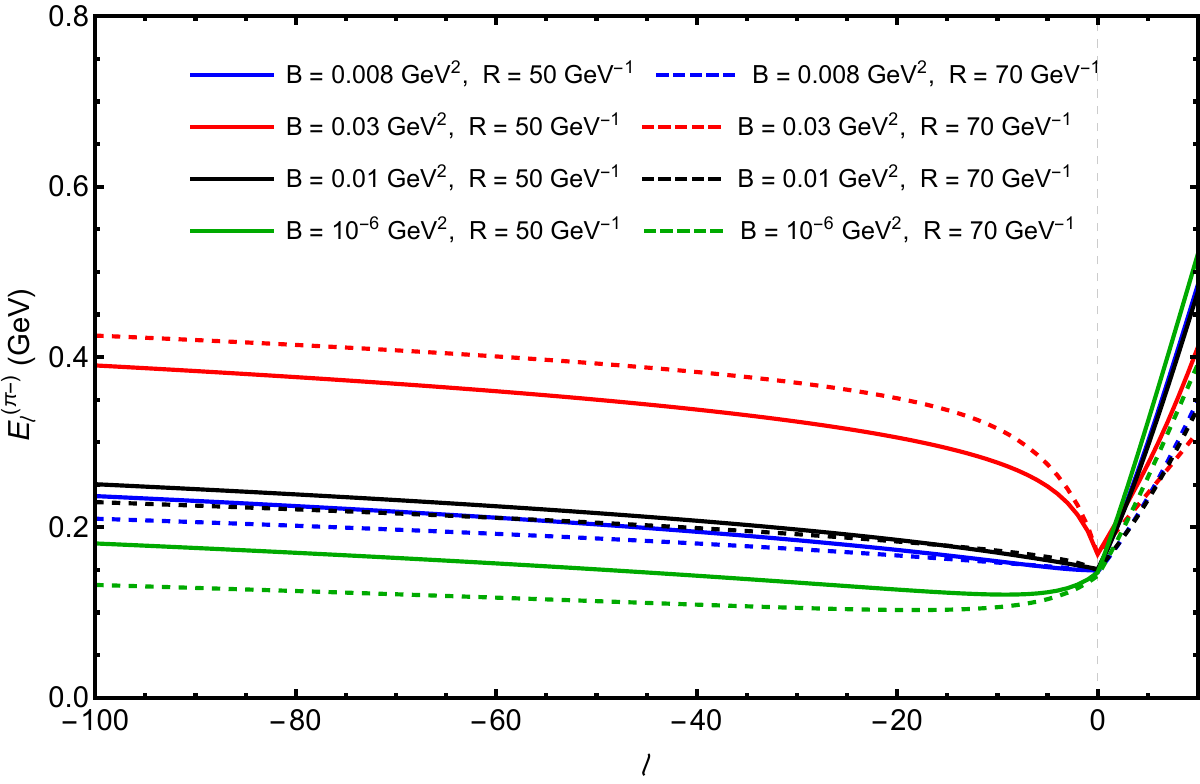}
    \caption{\justifying \small The variation of minimum energy for $\pi^-$ with $l$ for various values of $B$ and $R$ at $\Omega = 1/R$.}
    \label{Emin1}
\end{figure}
It is clear from the energy expression that for $\pi^+$, $l^* \ge 0$. We have plotted these quantities numerically in Figs.~\ref{omegac}, \ref{dotted}, and \ref{rcurve}. Fig.~\ref{omegac} shows the threshold chemical potential required for condensation to occur with $\Omega$ for different values of the magnetic field. We have shown the parameter range over which we perform our transport calculations. We see that there is a small range of $\Omega$ around $\Omega\sim0.005\,\,\text{GeV}$, where all the curves intersect. For the values of $\Omega$ smaller than this, with the increase in the magnitude of the magnetic field, the threshold chemical potential $\mu_c$ and rotation velocity $\Omega_c$ required for condensation increase, while for values of $\Omega>0.005\,\,\text{GeV}$, $\mu_c$ and $\Omega_c$ decrease with the increase in $B$. In Fig.~\ref{dotted}, we show the states that condense for a given magnetic field. Fig.~\ref{dotted1} shows this for the minimum $B$ that we have considered for our transport calculations ($B=0.008\,\,\text{GeV}^2$). Here we see that the lowest-lying state that condenses, $l^*$, increases starting from $0$, as the rotation velocity $\Omega$ increases. For $B=0.008\,\,\text{GeV}^2$, we see that the rise in $l^*$ with $\Omega$ is slower as compared to $B=0.03\,\,\text{GeV}^2$ case in Fig.~\ref{dotted2}. The effect of the finite radius of the system on $\mu_c$ and $\Omega_c$ is shown in Fig.~\ref{rcurve}. The increase in the size of the system decreases the $\mu_c$ and $\Omega_c$ required for condensation. We see that the effect of system size is more pronounced at values of $\Omega R$ close to the causality limit as compared to smaller values of $\Omega R$.

For $\pi^-$, the condition for condensation reads
\begin{align}\label{Elpi-}
    \sqrt{m^2+eB(2a_1(l)+|l|-l+1)}+\Omega l=0   
\end{align}
In the above, we have dropped the chemical-potential term, since it is positive and thus disfavors condensation, as discussed earlier in this section. For $l\ge 0$, both terms in Eq. \eqref{Elpi-} are positive, so condensation is not possible. Therefore, the only case that we need to study is $l<0$ in Eq.~\eqref{Elpi-}. To study this numerically, we plot minimum energy at a given $l$, $E_{l}=\sqrt{m^2+eB(2a_1(l)-2l+1)}+\Omega l$, with $l$, for $\Omega=1/R$ i.e. exactly at the causal boundary, as has been shown in Fig.~\ref{Emin1}. In the same figure, we see that for the minimum and maximum magnetic field that we work with in the next sections ($B=0.008\,\,\text{GeV}^2$ and $B=0.03\,\,\text{GeV}^2$ respectively), the minima of $E_l$ are attained at $l^*=0$ and $E_{l^*}>0$, which implies that in our working range of magnetic field, condensation does not occur even at the causal boundary of $\Omega$. We also note that the minima of $E_l$ decrease as the magnetic field decreases, and the line towards negative $l$ looks almost parallel to the x-axis for our minimum magnetic field. Thus, in the same figure, we also consider a small value of magnetic field $B = 10^{-6}\,\,\text{GeV}^2$ to verify whether the system condenses for such small magnetic fields. We see that while the $l$ at which minima are attained become more negative, even at such small values of the magnetic field, $E_{l^*}$ remains positive, and $\pi^-$ do not condense. Therefore, while the infinite-volume analysis \cite{zahed} yields symmetric condensation thresholds, introducing a finite system radius reveals that $\pi^+$ and $\pi^-$ no longer require identical conditions to condense.

\section{Transport Coefficients}\label{TC}
We now turn to the evaluation of the transport coefficients for the rotating, magnetized pion gas. Utilizing the BTE under the RTA, we derive the formal expressions for the longitudinal electrical conductivity, thermal conductivity, and the Seebeck coefficient. Furthermore, we define a normalized Lorenz ratio to systematically probe the relative efficiencies of heat and charge transport. In this section, as in Ref. \cite{padhan}, we assume that $\Omega x$, $\Omega y$, and $\Omega/T$ are sufficiently small, so that we work only with terms linear in them. Therefore, it should be understood that all equalities herein are valid up to this linear order. Thus, in LRF $u^{\mu}=(1,0,0,0)$. The equilibrium distribution functions for the charged and neutral pions then take the respective forms
\begin{align}\label{f0charged}
    f^0_j=\frac{1}{e^{\beta(p_{oj}-j\mu)}-1},
\end{align}
and
\begin{align}\label{f0n}
    f^0_u=\frac{1}{e^{\beta p_{0u}}-1}.
\end{align}
The invariant phase space measure for charged pions is modified as
\begin{align}
    \int \frac{d^3p}{(2\pi)^3p_{0j}}\rightarrow \frac{1}{A}\sum_{l=-\infty}^{\infty}\sum_{k=1}^{\infty}\int_{-\infty}^{\infty}\frac{dp_z}{(2\pi)p_{0j}} \equiv \sumint \frac{d\chi}{p_{0j}}
\end{align}
Here $p^3=-p_3=p_z$. For the case of neutral pions, the phase space measure remains structurally identical; however, the index $k$ labels the root of the Bessel function rather than the confluent hypergeometric function, and $p_{0j}$ is replaced by $p_{0u}$. Although the summations over $l$ and $k$ extend to infinity, all evaluated physical quantities are strictly convergent due to the exponential suppression at high energies introduced by the distribution function.
Therefore, to evaluate the transport coefficients numerically, we truncate the infinite summations by introducing upper limits $l_{\text{max}}$ and $k_{\text{max}}$ for the charged pions, and corresponding limits $l^u_{\text{max}}$ and $l^u_{\text{max}}$ for the neutral pions. 

Following Ref.~\cite{zahed}, we take the system radius to be $R = 50~\text{GeV}^{-1} \approx 10~\text{fm}$. To remain safely outside the condensation regime discussed in Sec.~\ref{PC}, we restrict our transport analysis to $\Omega \leq 0.006~\text{GeV}$, which keeps the system in the normal phase even for the strongest magnetic field we consider. For comparision \cite{zahed} used $0.005<\Omega<0.0085\,\,\text{GeV}$. Regarding the magnetic field, although the spectator nucleons in non-central collisions generate fields of order $B \sim m_{\pi}^2/e \approx 0.06~\text{GeV}^2$, these fields are transient. While the finite electrical conductivity of the medium extends their lifetime \cite{PhysRevC.107.034901,PhysRevC.82.034904,https://doi.org/10.1155/2013/490495,STEWART2021122308}, the field decays considerably before the pions are produced in the hadronization phase. We therefore adopt $B = 0.03~\text{GeV}^2$ as an upper estimate for the magnetic field present in the pionic medium.
   
\subsection{Electrical conductivity}\label{EC}
Only the charged pions participate in the electric response. The BTE for charged pions in a rotating frame is given by \cite{padhan}: 
\begin{equation}\label{BTE}
    p^\mu \frac{\partial f_j}{\partial x^\mu} - \Gamma^\alpha_{\mu \lambda} p^\mu p^\lambda \frac{\partial f_j}{\partial p^\alpha} + q_j F^{\alpha \beta} p_\beta \frac{\partial f_j}{\partial p^{\alpha}} = C[f]
\end{equation}
where $\Gamma^{\alpha}_{\mu\lambda}=\frac{1}{2}g^{\alpha\nu}\big(\partial_{\lambda}g_{\nu\mu}+\partial_{\mu}g_{\lambda\nu}-\partial_{\nu}g_{\mu\lambda}\big)$ are the connection coefficients. For the metric in Eq. \eqref{metric}, the only non-zero connection coefficients are:
\begin{equation}
\begin{split}
    \Gamma^1_{00} = -x \Omega^2, \quad \Gamma^2_{00} = -y \Omega^2, \\ \Gamma^1_{20} =\Gamma^1_{02} = - \Omega, \quad \Gamma^2_{10} =\Gamma^2_{01} =  \Omega
\end{split}
\end{equation}
$F^{\alpha\beta}$ is the electromagnetic field tensor. In this work, we set the magnetic field to zero in the force term of the BTE, since its effects are already fully incorporated into both the particle spectra and the equilibrium distribution function. In Eq. \eqref{BTE}, $f=f^0+\delta f$ where $f^0 $ is the equilibrium distribution function, $\delta f$ is the shift of the distribution away from the equilibrium due to the electric field, and is first order in derivatives. Keeping terms only up to first-order in derivatives and assuming (only for the case of electrical conductivity) that $f^0$ is independent of the space-time coordinate $x$, we get:
\begin{equation}
     - \Gamma^\alpha_{\mu \lambda} p^\mu p^\lambda \frac{\partial f^0_j}{\partial p^\alpha} + q F^{\alpha \beta} p_\beta \frac{\partial f^0_j}{\partial p^{\alpha}} = C[f].
\end{equation}
In LRF, the terms with $\Gamma^1_{00}$ and $\Gamma^2_{00}$ can be neglected as they are quadratic in $\Omega x$, $\Omega y$, and $\Omega/T$, while other terms with the connection coefficients cancel out. 

Since the dynamics of the charged particles are quantized in the $xy$-plane due to the magnetic field, we study the transport coefficients only in the direction of the magnetic field. Therefore, we assume the electric field to be along the z-direction, i.e. $\mathbf{E}=E_z\hat{\mathbf{z}}$. Retaining terms up to linear order in $\Omega x$, $\Omega y$, and $\frac{\Omega}{T}$, and employing RTA for the collision term, we get:
\begin{equation}
    q_j E_z p_{0j} \frac{\partial f^0_j}{\partial p_z} = C[f] = -p_{0j} \frac{\delta f_j}{\tau_R}.
\end{equation}
where $\tau_R$ is the relaxation time, which has been calculated in Ref.~\cite{kalikotay} by considering a $2\rightarrow2$ pion-pion scattering process with interactions occurring through the exchange of vector meson $\rho$ and scalar meson $\sigma$ in a thermal background, and fitting it with a polynomial function given by
\begin{equation*}
    \tau_R=\Sigma_{i=0}^{3}a_i\Big(\frac{m}{T}\Big)^i\frac{1}{T^3}
\end{equation*}
where $a_0=0.0145\,\text{fm}\,\text{GeV}^3$, $a_1=-0.0109\,\text{fm}\,\text{GeV}^3$, $a_2=0.0058\,\text{fm}\,\text{GeV}^3$, and $a_3=0.0026\,\text{fm}\,\text{GeV}^3$. This relaxation time does not include the effects of the magnetic field and rotation. We argue that this is still a good approximation to work with in our case because the interactions of pions are local and are not affected by the global rotation of the system. In Ref.~\cite{kalikotay2024electrical}, the authors have studied the behavior of the relaxation time for a $2\rightarrow2$ pion scattering process in the presence of a background magnetic field. Their results indicate that the relaxation time is approximately independent of the magnetic field. 

Using the expressions of equilibrium distribution functions for the charged pions, Eq. \eqref{f0charged}, we obtain,
\begin{equation}
    \delta f_j = q_j \tau_R E_z \frac{p_z}{p^0_{j}} \frac{f^0_j(1+f^0_j)}{T}
\end{equation}
where $p^0=p_0+j\Omega l$. The electric current induced in the medium due to the electric field is defined as
\begin{equation}\label{current}
    j_z=\sum_j\sumint \frac{d\chi}{p_{0j}}q_j\,p_{z}\,\delta f_j.
\end{equation}
Substituting $\delta f_j$, we get,
\begin{align}
    j_z = \frac{q^2 \tau_R E_z}{ T}\sum_j \sumint d\chi \frac{p_z^2}{p_{0j}p^0_j}  f_j^0(1+f_j^0)
\end{align}
Then, the longitudinal ohmic conductivity is,
\begin{align}\label{sigma}
    \sigma_{||} = \frac{q^2 \tau_R}{ T} \sum\limits_j\sumint d\chi \frac{p_z^2}{p_{0j}p^0_j}  f_j^0(1+f_j^0) 
\end{align}

\subsection{Thermal Conductivity}\label{ThermCond}
In the thermal response, all pions (charged and neutral) contribute, and for any system, it is characterized by the thermal conductivity. It is studied using the dissipative heat flow in the system, which in covariant form is defined for a species as
\begin{align}
    I_j^{\mu}=u_{\nu}\delta T_j^{\nu\sigma}\Delta^{\mu}_{\sigma}-h_j\delta N_{\sigma j}\Delta^{\mu}_{\sigma}
\end{align}
where $\delta T^{\mu\nu}$ and $\delta N^{\mu}$ are the non-equilibrium parts of the energy-momentum tensor and particle four-flow, respectively, and are defined as
\begin{align}
    \delta T^{\mu\nu}_j&=\sumint \frac{d\chi}{p_{0j}}p^{\mu}_jp_{j}^{\nu}\delta f_j \label{deltaT}\\
    \delta N^{\mu}_j&=\sumint \frac{d\chi}{p_{0j}}p_{j}^{\mu}\delta f_j \label{deltaN}
\end{align}
with $\delta f_j$ being the shift of the distribution away from equilibrium due to temperature gradients in the system. $\Delta^{\mu\nu}=g^{\mu\nu}-u^{\mu}u^{\nu}$ is the projection operator orthogonal to the four-velocity $u^{\mu}$, and $h_j=\frac{\varepsilon_j+P_j}{n_j}$ is the enthalpy per particle of the $j^{\text{th}}$ species with $\varepsilon_j$, $P_j$, and $n_j$ being the energy density, pressure, and number density, respectively, of that species. They are defined as
\begin{align}
    \varepsilon_j&=u_{\mu}u_{\nu}T^{\mu\nu}_{0j},\label{ed}\\ 
    P_j&=-\frac{1}{3}\Delta_{\mu\nu}T^{\mu\nu}_{0j},\label{P} \\ 
    n_j&=u_{\mu}N^{\mu}_{0j}.\label{n}
\end{align}

We start by calculating the contribution of the charged pions. In LRF, the expressions for energy density, pressure, and number density can be derived using the above equations. They take the following forms:
\begin{align}
    \epsilon_j &=  \sumint d \chi \,p_{0j}\Bigg[1-\Big(\frac{j\Omega l}{p_{0j}}\Big)^2\Bigg]f^0_j, \\
    P_j&=\frac{1}{3 } \sumint d\chi \Bigg[\frac{(p_{0j}+j\Omega l)^2-m^2}{p_{0j}}\Bigg]f^0_j, \\
    n_j &=  \sumint d\chi \,f_j^0,
\end{align}
The longitudinal component of the heat flow in LRF simplifies to:
\begin{align}\label{Iparallel}
    I_\parallel^{\text{charged}}&=\sum_j(g_{0\beta}\,\delta T^{3\beta}_j-h_j\delta N^{3}_j) \\\nonumber
    &= \sum_{j}(\delta T^{30}_j+\Omega y \,\delta T^{31}_j-\Omega x\, \delta T^{32}_j-h_j \delta N_j^3)
\end{align}
Using Eq. \eqref{deltaT} and Eq. \eqref{deltaN} in Eq. \eqref{Iparallel}, we get
\begin{align}\label{I||}
    I_\parallel^{\text{charged}} = \sum \limits_{j} \sumint \frac{d\chi}{p_{0j}} p_{zj} (p_{0j} - h_j) \delta f_j.
\end{align}
To determine the shift $\delta f_j$, we employ the BTE: 
\begin{equation}
    p^\mu \frac{\partial f_j}{\partial x^\mu} - \Gamma^\alpha_{\mu \lambda} p^\mu p^\lambda \frac{\partial f_j}{\partial p^\alpha}  = -(u.p) \frac{\delta f_j}{\tau_R},
\end{equation}
Assuming that the gradients of temperature and chemical potential are only along the z-direction, in LRF, this simplifies to
\begin{equation}\label{deltaf}
    \delta f_j = -\frac{\tau_R}{p_0} p_{zj} \frac{\partial f^0_j}{\partial z}.
\end{equation}
Using the equilibrium distribution function from Eq. \eqref{f0charged}, we get
\begin{align}\label{df0/dz}
    \frac{\partial f^0_j}{\partial z}=\frac{f^0_j(1+f^0_j)}{T}\Big\{\frac{p_{0j}-j\mu}{T}\frac{dT}{dz}+j\frac{d\mu}{dz}\Big\}.
\end{align}
We want to relate the heat flow with the temperature gradient $\frac{dT}{dz}$. To do this, we use the Gibbs-Duhem relation, which in covariant form is \cite{Jaiswal2016,ISRAEL1979341,ISRAEL1976310}
\begin{align}
    u_{\mu}\big[d(P_j\beta^{\mu})-N_{0j}^{\mu}\,d\alpha_j+T_{0j}^{\mu\nu}d\beta_{\nu}\big]=0,
\end{align}
where $\beta^{\mu}=\frac{u^{\mu}}{T}$, and $\alpha_j=j\frac{\mu}{T}$. In LRF, using Eqs. \eqref{ed}, \eqref{P}, and \eqref{n}, we get
\begin{align*}
    d\mu=\frac{1}{n_j}dP_j-\frac{(h_j-j\mu)}{T}dT.
\end{align*}
Imposing mechanical equilibrium, i.e. $dP=0$, and using this in Eq. \eqref{df0/dz}, gives
\begin{align}\label{df/dz}
    \frac{\partial f^0_j}{\partial z} = \frac{f^0_j(1+f^0_j)}{T^2} &\big(p_{0j}-h_j\big)\frac{dT}{dz}.
\end{align}
Using Eqs. \eqref{df/dz} and \eqref{deltaf} in Eq. \eqref{I||}, we get
\begin{align}
    I_\parallel^{\text{charged}}=-\frac{\tau_R}{T^2}\sum_j\sumint d\chi \Big(\frac{p_z}{p_{0j}}\Big)^2 &(p_{0j}-h_j)^2 \\ \nonumber &\times f^0_j(1+f^0_j)\frac{dT}{dz}
\end{align}
Since $I_\parallel^{\text{charged}}=-\kappa_\parallel^{\text{charged}}\frac{dT}{dz}$, we get
\begin{align}\label{chargedkappa}
    \kappa_\parallel^{\text{charged}}=\frac{\tau_R}{T^2}\sum_j\sumint d\chi\Big(\frac{p_z}{p_{0j}}\Big)^2  (p_{0j}&-h_j)^2 \\ \nonumber &\times f^0_j(1+f^0_j).
\end{align}

For the case of neutral pions, since they are electrically neutral, they do not contribute to the charge current $N^{\mu}$, which is the only conserved current in the system. Therefore, the heat flow for the case of neutral pions is
\begin{align}
    I_\parallel^{\text{neutral}}= \delta T^{30}+y\Omega\,\delta T^{31}-x\Omega\,\delta T^{32}
\end{align}
Using Eqs. \eqref{deltaT} and \eqref{deltaN}, we get
\begin{align}\label{I||n}
    I_\parallel^{\text{neutral}}=\sumint d \chi p_z\delta f_u
\end{align}
where the shift $\delta f$ is of the same form as in Eq. \eqref{deltaf}, but instead of the $f^0_j$, we have $f^0_u$ of Eq. \eqref{f0n}, using which one can evaluate $\frac{\partial f^0_u}{\partial z}$, and hence $\delta f_u$. Using this in Eq. \eqref{I||n}, we can obtain the thermal conductivity contribution of neutral pions as
\begin{align}
    \kappa_\parallel^{\text{neutral}}=\frac{\tau_R}{T^2}\sumint d\chi \,\,p_z^2f^0_u(1+f^0_u).
\end{align}
The total thermal conductivity is then obtained as the sum of the contributions from the charged and neutral pions:
\begin{align}
    \kappa_\parallel=\kappa_\parallel^{\text{charged}}+\kappa_\parallel^{\text{neutral}}.
\end{align}
We discuss the results in section \ref{RD}.

\subsection{Seebeck Coefficient}\label{SC}
The temperature gradient in a system of charged particles can cause the charges to separate and induce an electric field in the medium. This effect is called the Seebeck effect, and for any medium it is characterized by the Seebeck coefficient, $S$, defined as $S=\frac{E_z}{dT/dz}$. We follow the methodology of Ref. \cite{PhysRevD.102.014030} to obtain the Seebeck coefficient in our case. We start with the expression for the induced electric current in the medium due to the induced electric field given by Eq. \eqref{current}, but now the shift in the distribution away from equilibrium is caused by the thermal gradient and the induced electric field. In this case, the BTE becomes
\begin{align}
    p^{\mu}\frac{\partial f^0_j}{\partial x^{\mu}}-\Gamma^{\alpha}_{\mu\lambda}p^{\mu}p^{\lambda}\frac{\partial f^0_j}{\partial p^{\alpha}}+q_jF^{\alpha\beta}&p_{\beta}\frac{\partial f^0_j}{\partial p^{\alpha}}\\\nonumber
    &=-(u.p)\frac{\delta f_j}{\tau_R}.
\end{align}
Assuming again that the gradients of temperature and chemical potential are only along the z-direction, the induced electric field will then also be in the same direction. With this, the BTE simplifies to
\begin{align}
    p_z\frac{\partial f^0_j}{\partial z}+q_jE_zp_{0j}\frac{\partial f^0_j}{\partial p_z}=-p_{0j}\frac{\delta f_j}{\tau_R}.
\end{align}
Using Eq. \eqref{f0charged} and the Gibbs-Duhem relation as in the previous section, one can derive the shift $\delta f_j$ as
\begin{align}
    \delta f_j=-\tau_R\frac{p_z}{p_{0j}}\frac{f^0_j(1+f^0_j)}{T^2}\Big\{(p_{0j}&-h_j)\frac{dT}{dz} \\ \nonumber 
    &-q_jE_zT\frac{p_{0j}}{p^0_j}\Big\}.
\end{align}
The induced current can then be written as
\begin{align}
    j_\parallel=-\frac{\tau_R}{T^2}\sum_j\Big[(L_1)_j\frac{dT}{dz}-(L_2)_jE_z\Big]
\end{align}
where
\begin{align}
    (L_1)_j&=q_j\sumint d\chi \Big(\frac{p_z}{p_{0j}}\Big)^2(p_{0j}-h_j)f^0_j(1+f^0_j), \\
    (L_2)_j&=q_j^2T\sumint d\chi \frac{p_z^2}{p_{0j}p^0_j}f^0_j(1+f^0_j).
\end{align}
To obtain the Seebeck coefficient, $S$, we set $j_\parallel=0$, which makes the electric field proportional to the temperature gradient, and the proportionality factor is $S$. We get:
\begin{align}
    S=\frac{(L_1)_++(L_1)_-}{(L_2)_++(L_2)_-}.
\end{align}
We discuss our results in the next section.

\subsection{Relative significance of charge and thermal transport}\label{WFL}
The Wiedemann-Franz law states that the ratio of thermal conductivity to the electrical conductivity of a metal is directly proportional to the temperature. i.e. $\frac{\kappa}{\sigma}=LT$, where $L$ is the Lorenz number. It provides a fundamental probe into the relative significance of heat and charge transport within a medium. In the context of heavy-ion collisions, the Wiedemann-Franz law has been studied under various contexts \cite{PhysRevD.100.051503,dwibedi2025wiedemann,rath2019violation,PhysRevD.108.094007,PhysRevC.107.014910,PhysRevD.106.034008}. These studies conclude that the Wiedemann-Franz law is violated for the medium created in heavy-ion collisions. In this work, we focus on how rotation and magnetic field affect the relative dominance of charge and heat transport in our system via the Lorenz number. 

To systematically isolate the effects of the magnetic field and rotation on the Lorenz number, we define a normalized Lorenz ratio, $L/L_0$, where $L_0$ is the Lorenz number for the system without the influence of the magnetic field and rotation. To obtain $L_0$, we need to derive the expressions of $\sigma_{\parallel}$ and $\kappa_{\parallel}$ in the absence of the magnetic field and rotation. To do this, we note that in the absence of a magnetic field, the spectrum of charged pions becomes identical to that of neutral pions:
\begin{align}
    E_{l,k}(p_{\perp},p_z)=p_{0j}=\sqrt{m^2+p_z^2+\Big(\frac{\alpha_k(l)}{R}\Big)^2}-j\Omega l.
\end{align}
To switch off rotation, we simply set $\Omega=0$ in all the equations. Using this in Eqs.~\eqref{f0charged}, \eqref{sigma}, and \eqref{chargedkappa}, we can obtain $\sigma_{\parallel}$ and $\kappa_{\parallel}$ without the influence of the magnetic field and rotation, and hence, $L_0$ can be obtained. We discuss the results in Section \ref{RD}.

\section{Results and discussion}\label{RD}
Before discussing transport, we re-emphasize the central message of Section \ref{PC}. While infinite-volume approximations suggest that rotation linearly shifts Landau levels and triggers condensation, our analysis demonstrates that imposing physically realistic finite-radius boundary conditions significantly alters these thresholds. Specifically, within the computationally accessible range of magnetic field and system radii, which is bound by the numerical limits of obtaining the roots of confluent hypergeometric function $_1F_1$, we found that the lowest-lying energy states for $\pi^-$ remain strictly positive, therefore, even when pushed to the maximum causal angular velocity ($\Omega=1/R$), $\pi^-$ condensation is precluded. For $\pi^+$, although condensation is possible, the required threshold values for the charge chemical potential ($\mu_c$) and rotation ($\Omega_c$) fall well outside our physical operating region. Consequently, our chosen working regime is situated safely outside the condensation boundaries, ensuring that the transport phenomena evaluated herein are robust and unaffected by condensation-induced effects.
\begin{figure*}[ht]
    \centering
    \begin{subfigure}[b]{0.3\textwidth}
        \centering
        \includegraphics[width=\textwidth]{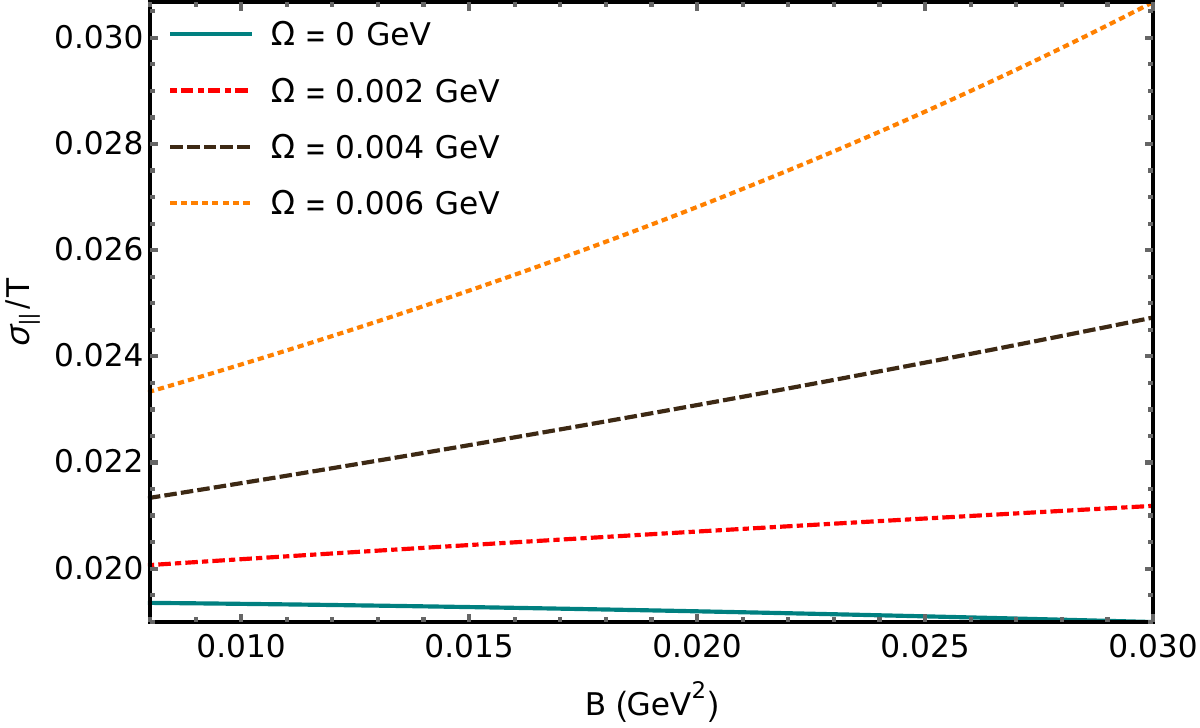}
        \caption{}
        \label{sigma/B}
    \end{subfigure}
    \begin{subfigure}[b]{0.3\textwidth}
        \centering
        \includegraphics[width=\textwidth]{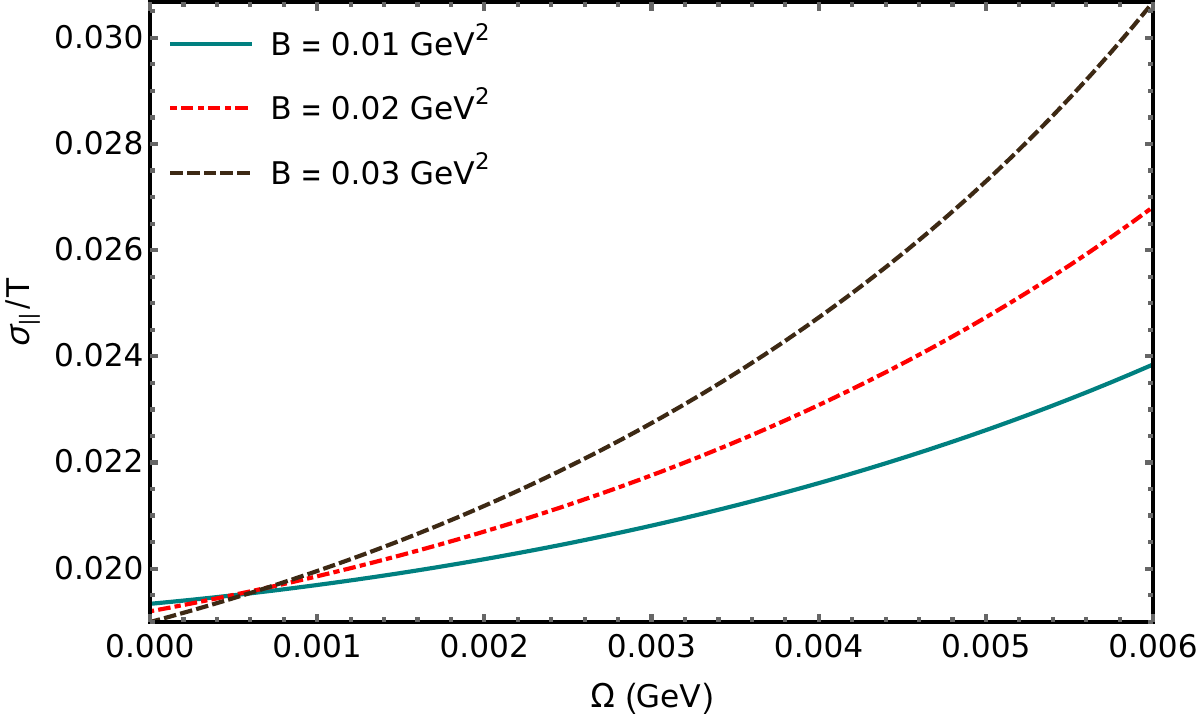}
        \caption{}
        \label{sigma/omega}    
    \end{subfigure}
    \begin{subfigure}[b]{0.3\textwidth}
        \centering
        \includegraphics[width=\textwidth]{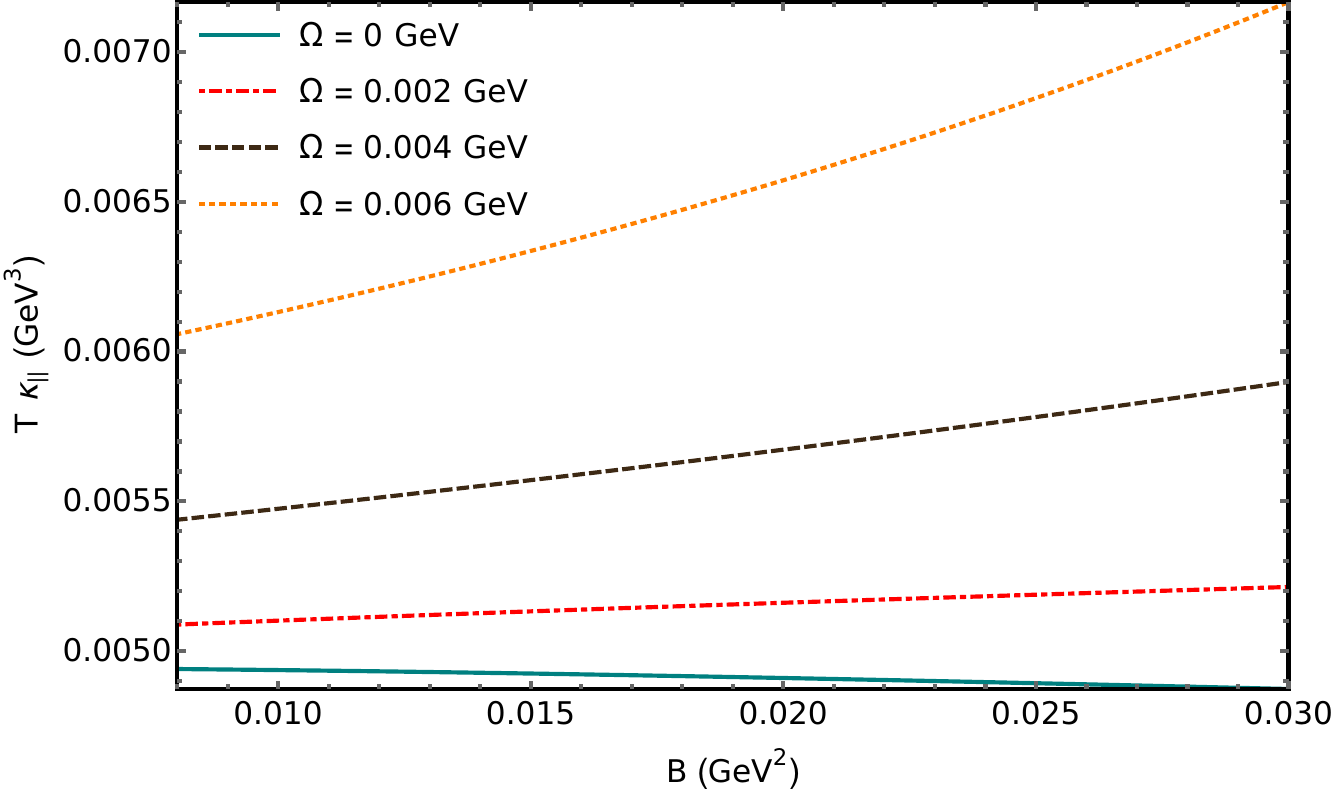}
        \caption{}
        \label{kappa/B}
    \end{subfigure}
    \hspace{0.5cm}
    \begin{subfigure}[b]{0.3\textwidth}
        \centering
        \includegraphics[width=\textwidth]{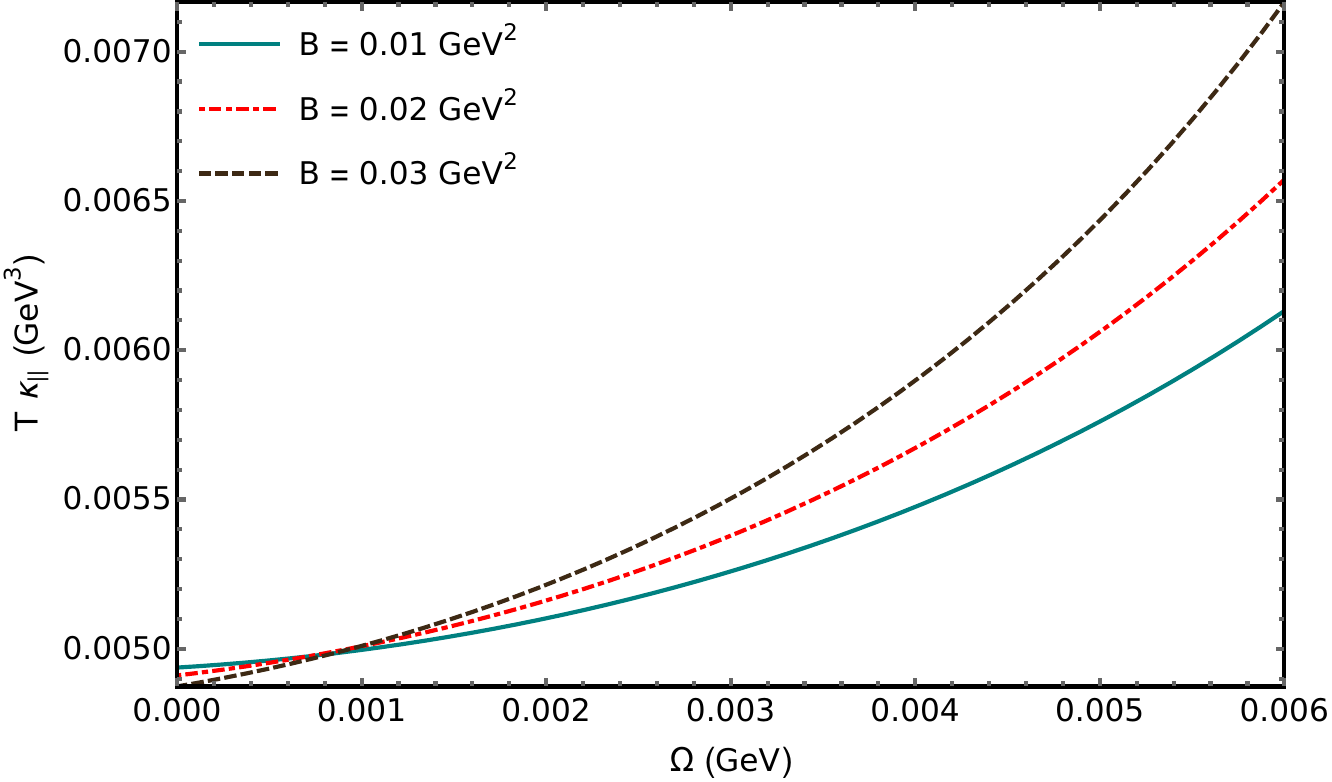}
        \caption{}
        \label{kappa/omega}    
    \end{subfigure}
    \begin{subfigure}[b]{0.3\textwidth}
        \centering
        \includegraphics[width=\textwidth]{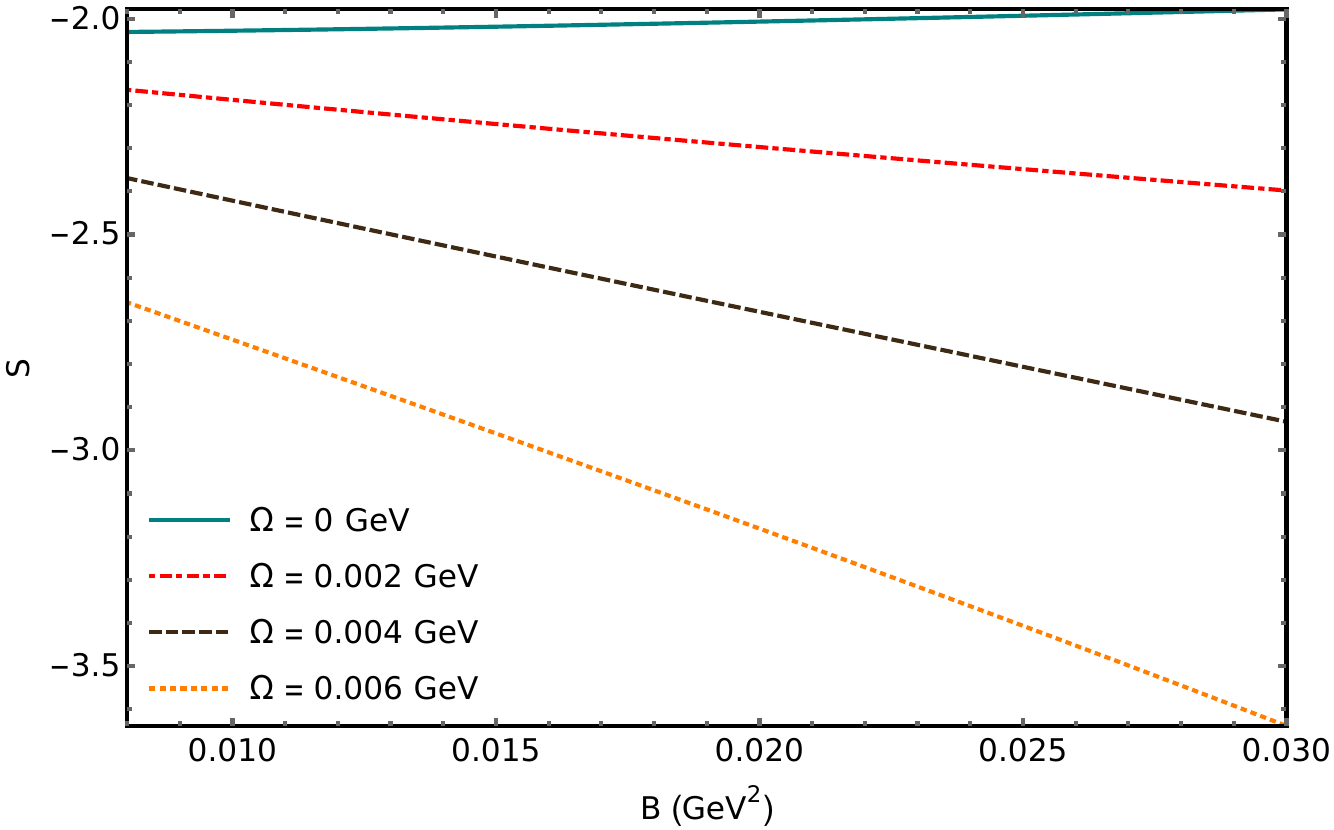}
        \caption{}
        \label{S/B}
    \end{subfigure}
    \begin{subfigure}[b]{0.3\textwidth}
        \centering
        \includegraphics[width=\textwidth]{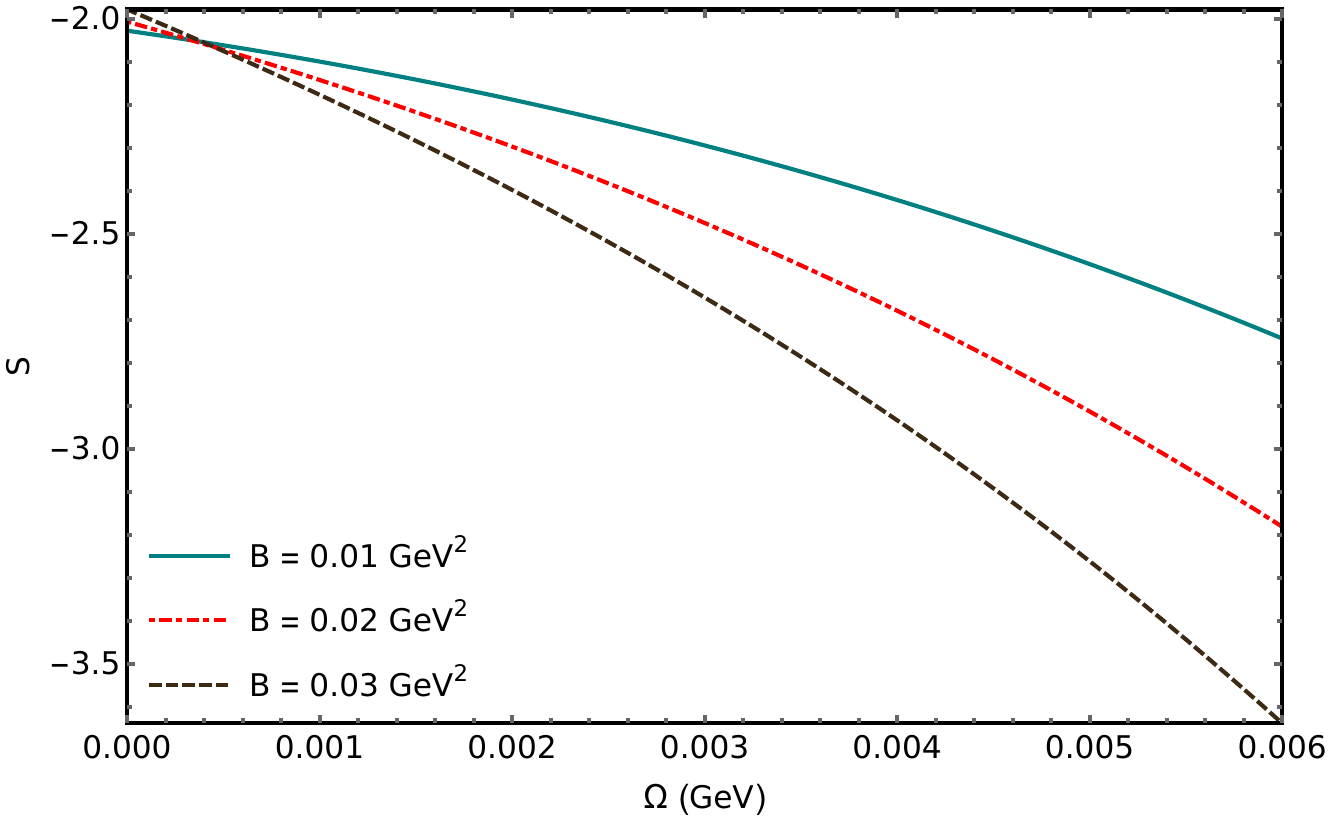}
        \caption{}
        \label{S/omega}    
    \end{subfigure}
    \caption{\justifying \small Transport coefficients with magnetic field and angular velocity of rotation for $R=50\,\,\text{GeV}^{-1}$ at $T=0.14\,\,\text{GeV}$ and $\mu=0.1\,\,\text{GeV}$. All the transport coefficients show a competing interplay between the magnetic field and rotation: while the magnetic field suppresses the transport coefficients in a static medium, rotation, acting as an effective chemical potential, introduces an energy shift that favors their increase. Beyond an angular velocity that varies with $B$, this rotational enhancement overpowers the magnetic suppression, leading to an increase in the transport coefficients with increasing magnetic field.}
    \label{TC/Bomega}
\end{figure*}
\begin{figure*}[ht]
    \centering
    \begin{subfigure}[b]{0.3\textwidth}
        \centering
        \includegraphics[width=\textwidth]{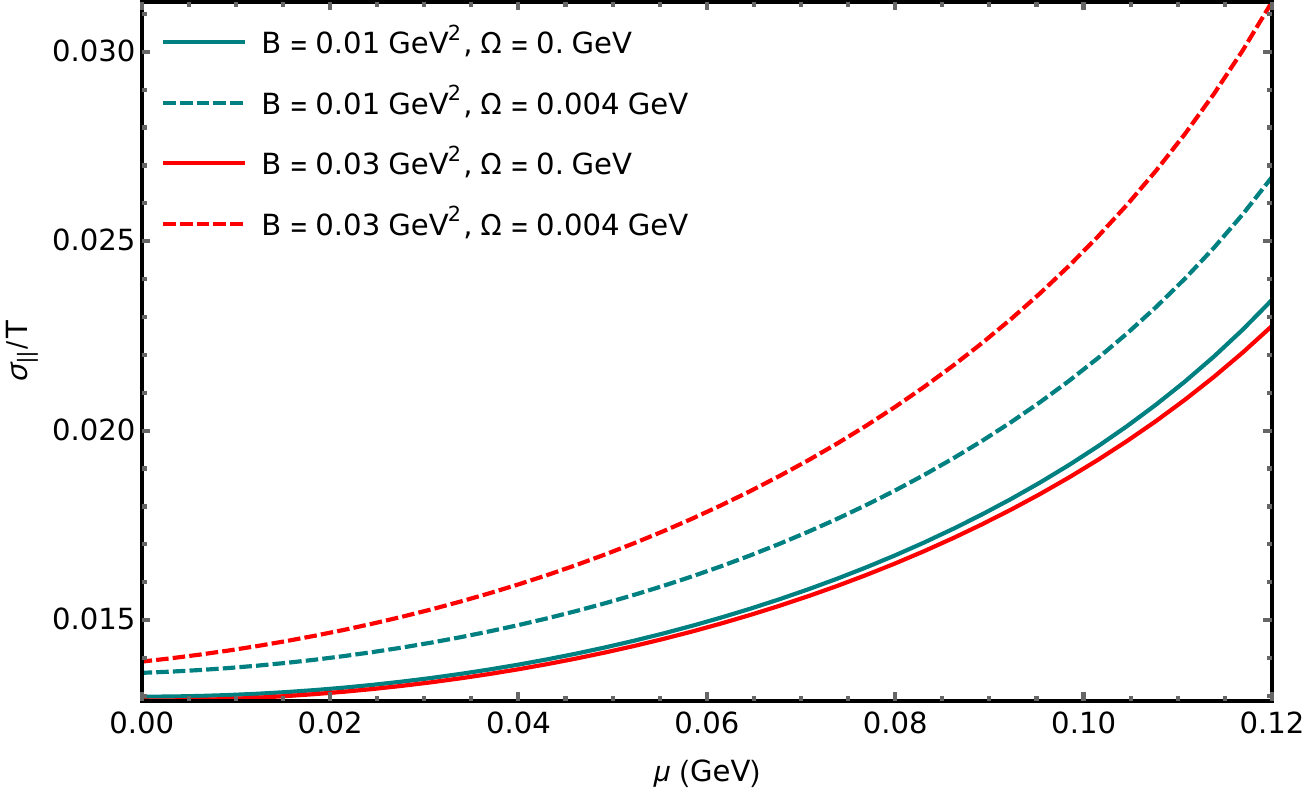}
        \caption{}
        \label{sigma/mu}
    \end{subfigure}
    \begin{subfigure}[b]{0.3\textwidth}
        \centering
        \includegraphics[width=\textwidth]{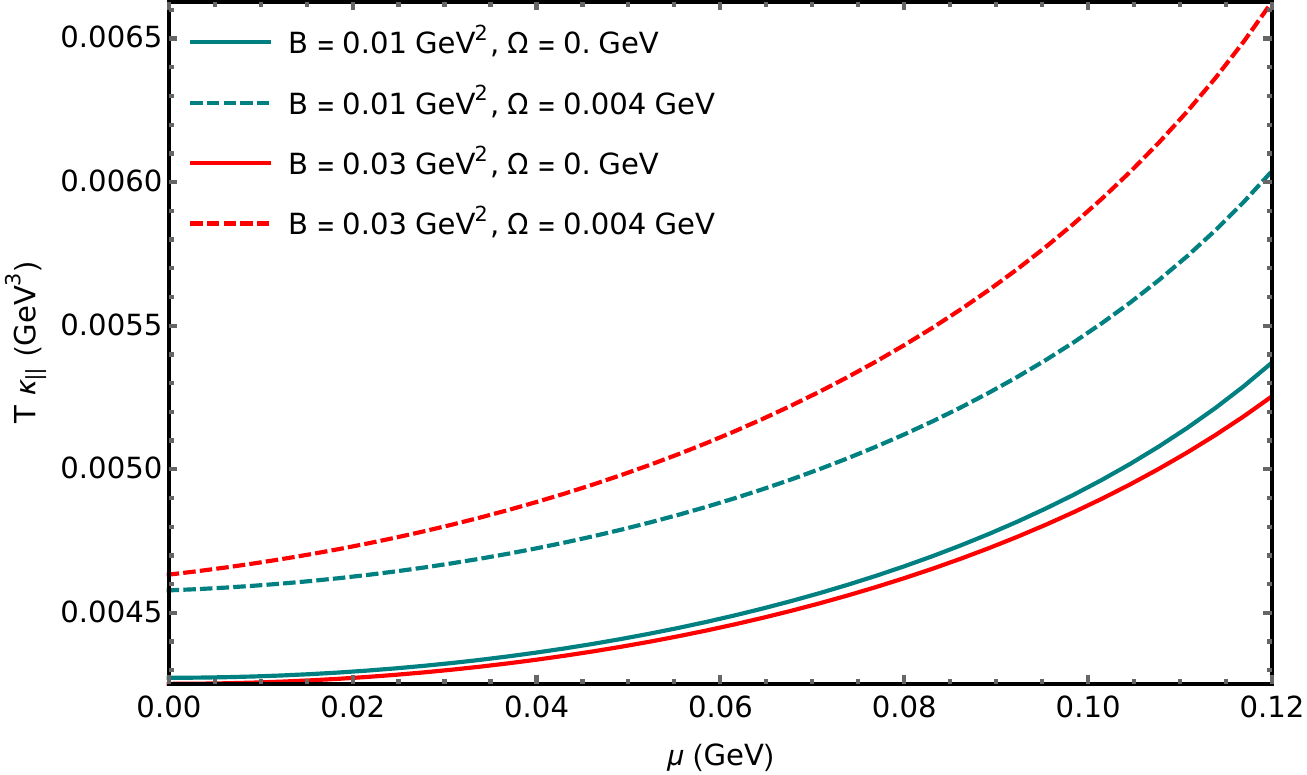}
        \caption{}
        \label{kappa/mu}    
    \end{subfigure}
    \begin{subfigure}[b]{0.29\textwidth}
        \centering
        \includegraphics[width=\textwidth]{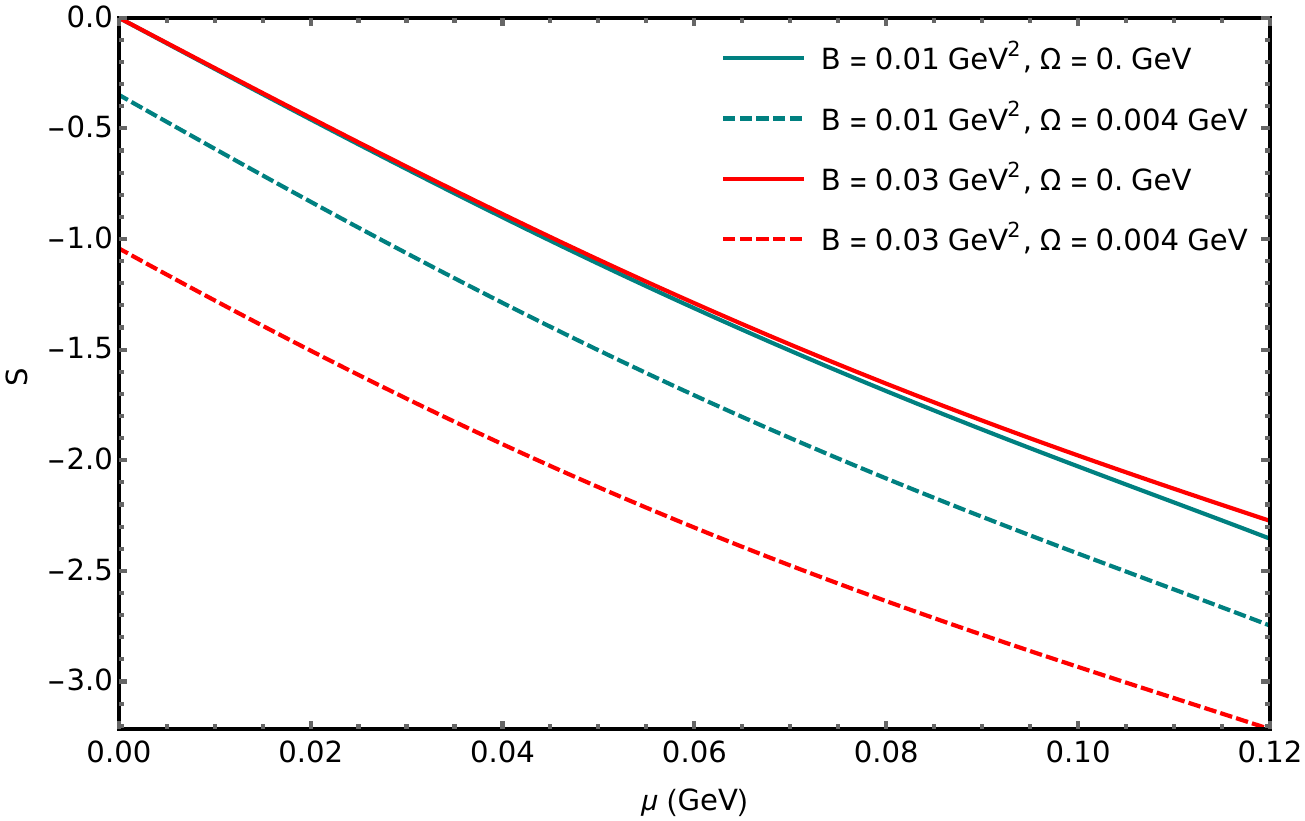}
        \caption{}
        \label{S/mu}
    \end{subfigure}
    \begin{subfigure}[b]{0.3\textwidth}
        \centering
        \includegraphics[width=\textwidth]{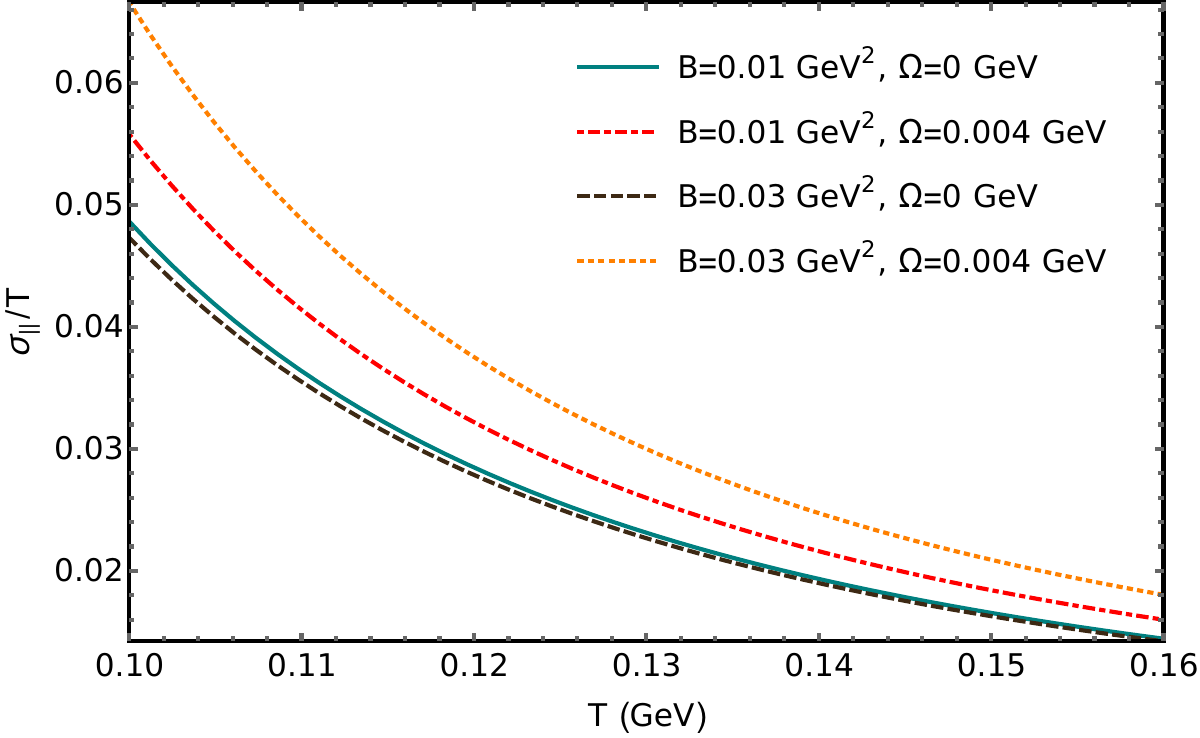}
        \caption{}
        \label{sigma/T}
    \end{subfigure}
    \begin{subfigure}[b]{0.3\textwidth}
        \centering
        \includegraphics[width=\textwidth]{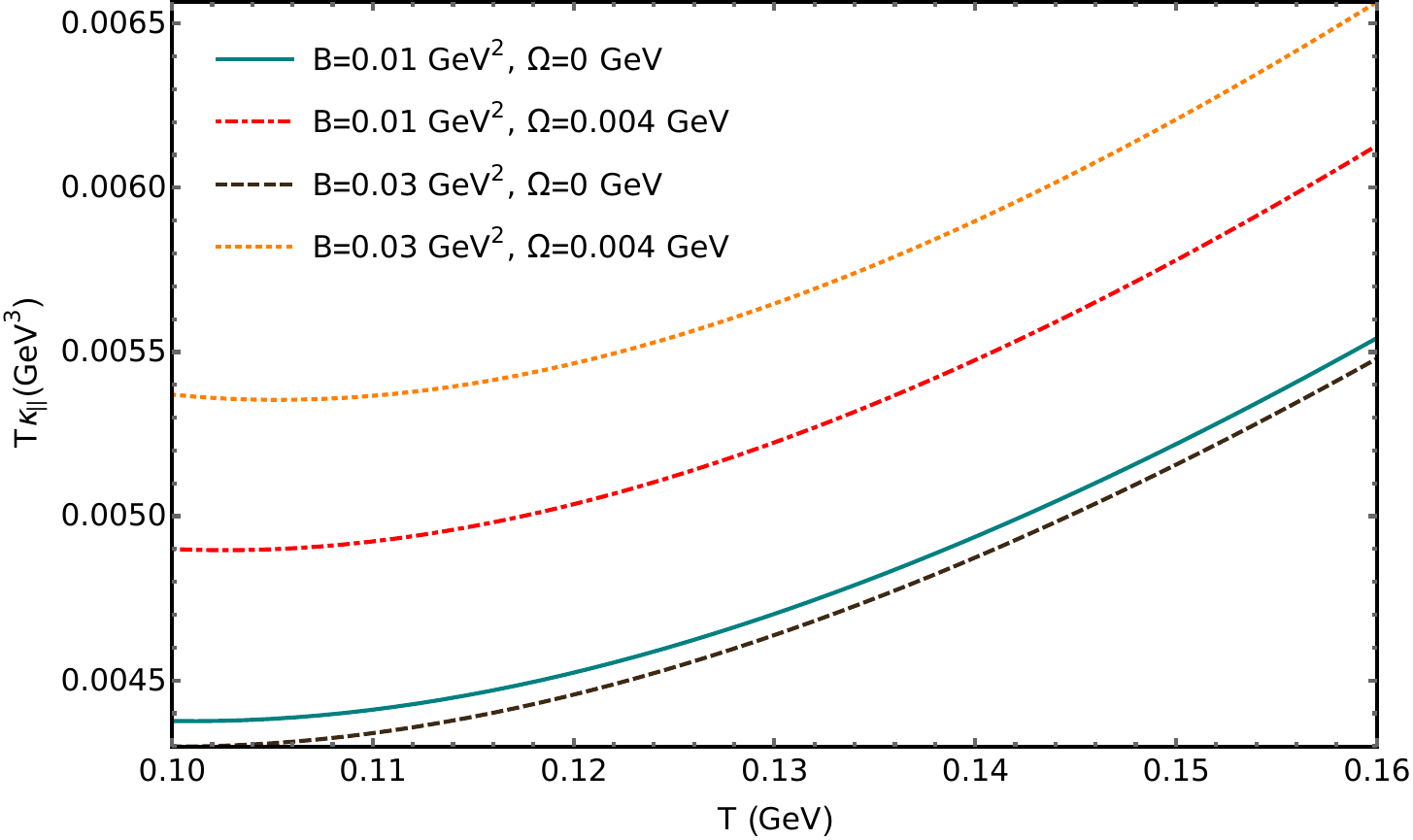}
        \caption{}
        \label{kappa/T}    
    \end{subfigure}
    \begin{subfigure}[b]{0.3\textwidth}
        \centering
        \includegraphics[width=\textwidth]{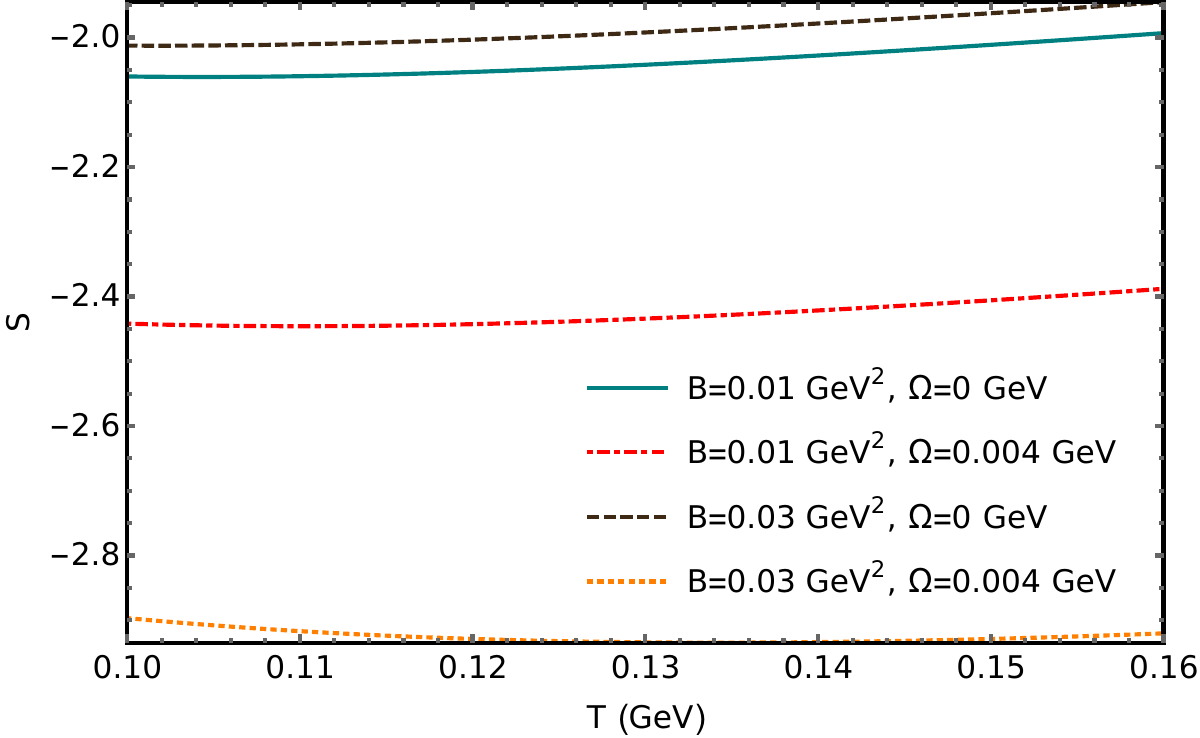}
        \caption{}
        \label{S/T}
    \end{subfigure}
    \caption{\justifying \small Transport coefficients with the charge chemical potential $\mu$ ($T=0.14\,\,\text{GeV}$) and temperature $T$ ($\mu=0.1\,\,\text{GeV}$) for $R=50\,\,\text{GeV}^{-1}$. The magnitudes of all the transport coefficients increase with $\mu$; notably, finite rotation amplifies the magnetic field's influence at higher chemical potentials and induces a non-zero Seebeck effect even at $\mu=0$ by acting as an effective chemical potential. Across all evaluated thermal and chemical regimes, the introduction of finite rotation consistently enhances the transport coefficients.}
\end{figure*}

Fig.~\ref{sigma/B} shows the variation of temperature-scaled longitudinal electrical conductivity $\sigma_{\parallel}/T$ with magnetic field for different angular velocities of rotation $\Omega$. We see that when $\Omega=0$, i.e., in the absence of rotation, $\sigma_{\parallel}/T$ decreases with the increase in the magnitude of the magnetic field. This is because, as $B$ increases, the energy of the charged pions, Eq.~\eqref{E}, increases, thereby suppressing the longitudinal phase space accessible to the charge carriers. This trend is consistent with the results of Ref.~\cite{PhysRevC.107.034903} for the temperature range we are working in. However, as $\Omega$ increases, this behavior reverses, and it starts increasing with $B$. The increase is almost linear, and as $\Omega$ increases, the rate of change in $\sigma_{\parallel}$ with $B$ also increases. We observe a similar trend in the behavior of thermal conductivity and Seebeck coefficient in Figs.~\ref{kappa/B} and \ref{S/B}. The variation of $\sigma_{\parallel}/T$ with $\Omega$ for different values of $B$ is shown in Fig.~\ref{sigma/omega}. Here, the $\Omega$ for which the change in the behavior of $\sigma_{\parallel}/T$ with $B$ inverts, is much clearer. Upon careful inspection, one can see that the points of intersection of the three lines are not the same, but are close. This is due to the fact that the behavior of $\sigma_{\parallel}/T$ with $B$ in Fig.~\ref{sigma/B} is not linear, but is close to linear. 

\begin{figure*}[ht]
    \centering
    \begin{subfigure}[b]{0.45\textwidth}
        \centering
        \includegraphics[width=\textwidth]{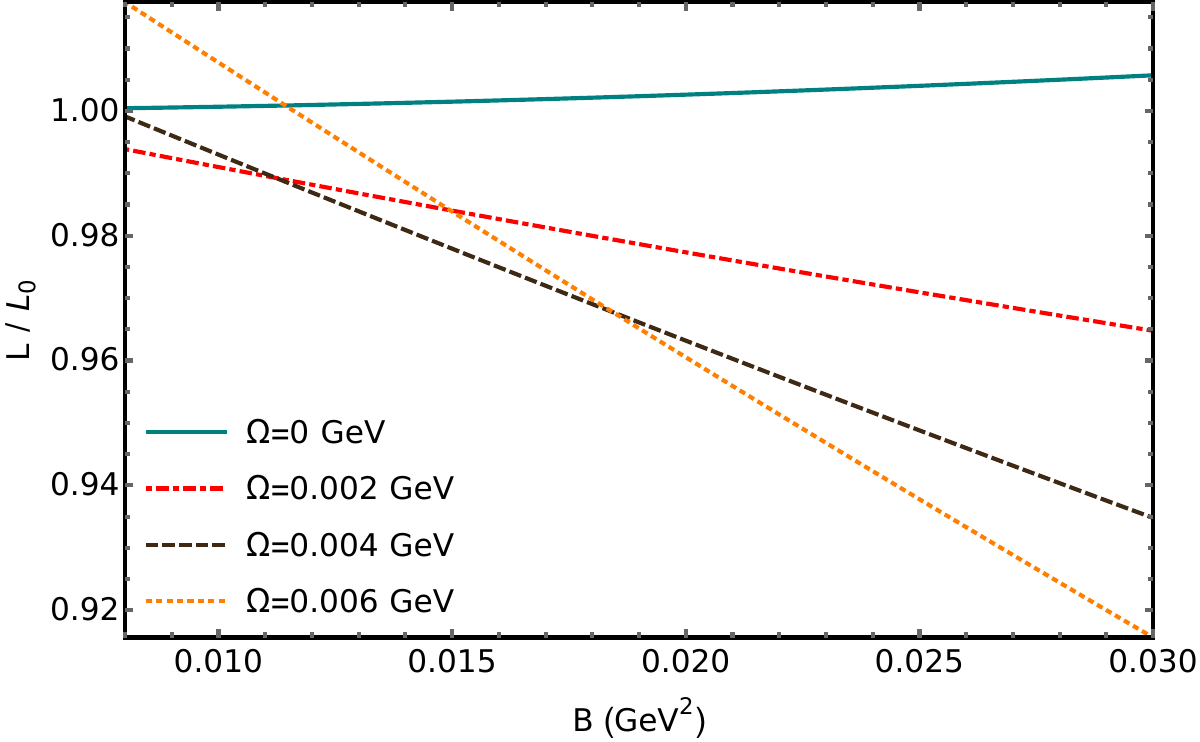}
        \caption{}
        \label{NormL/B}
    \end{subfigure}
    \begin{subfigure}[b]{0.45\textwidth}
        \centering
        \includegraphics[width=\textwidth]{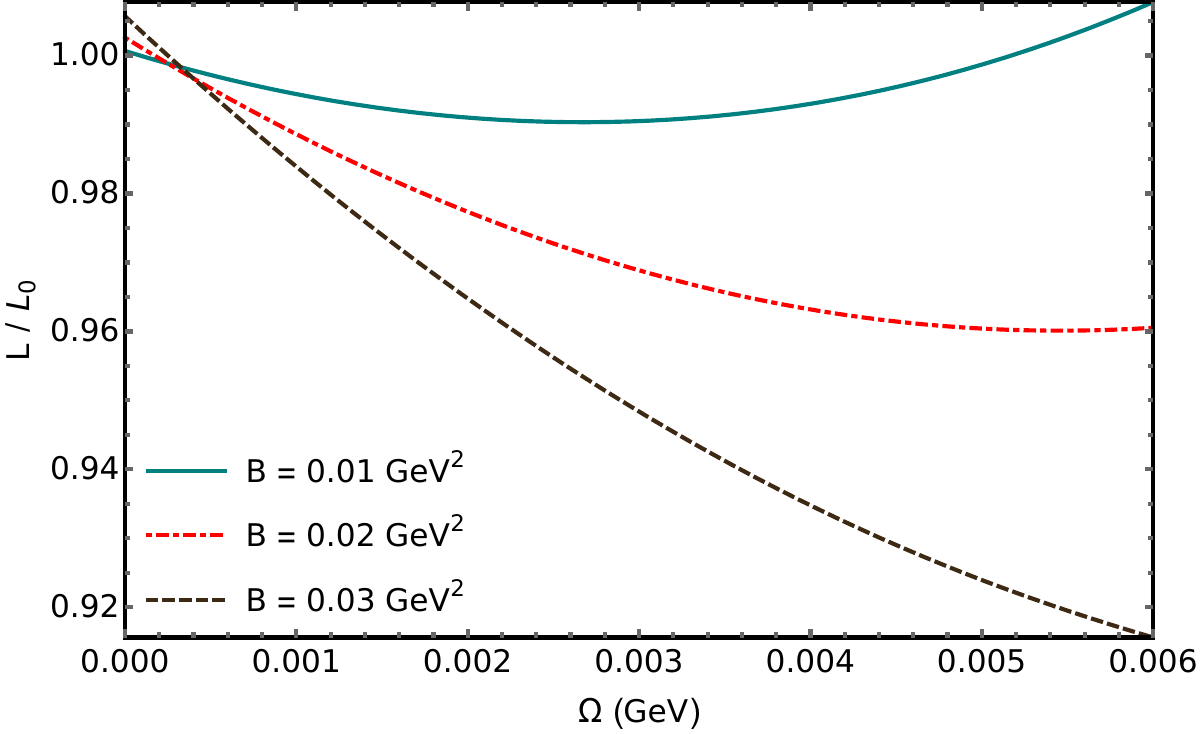}
        \caption{}
        \label{NormL/omega}    
    \end{subfigure}
    \begin{subfigure}[b]{0.45\textwidth}
        \centering
        \includegraphics[width=\textwidth]{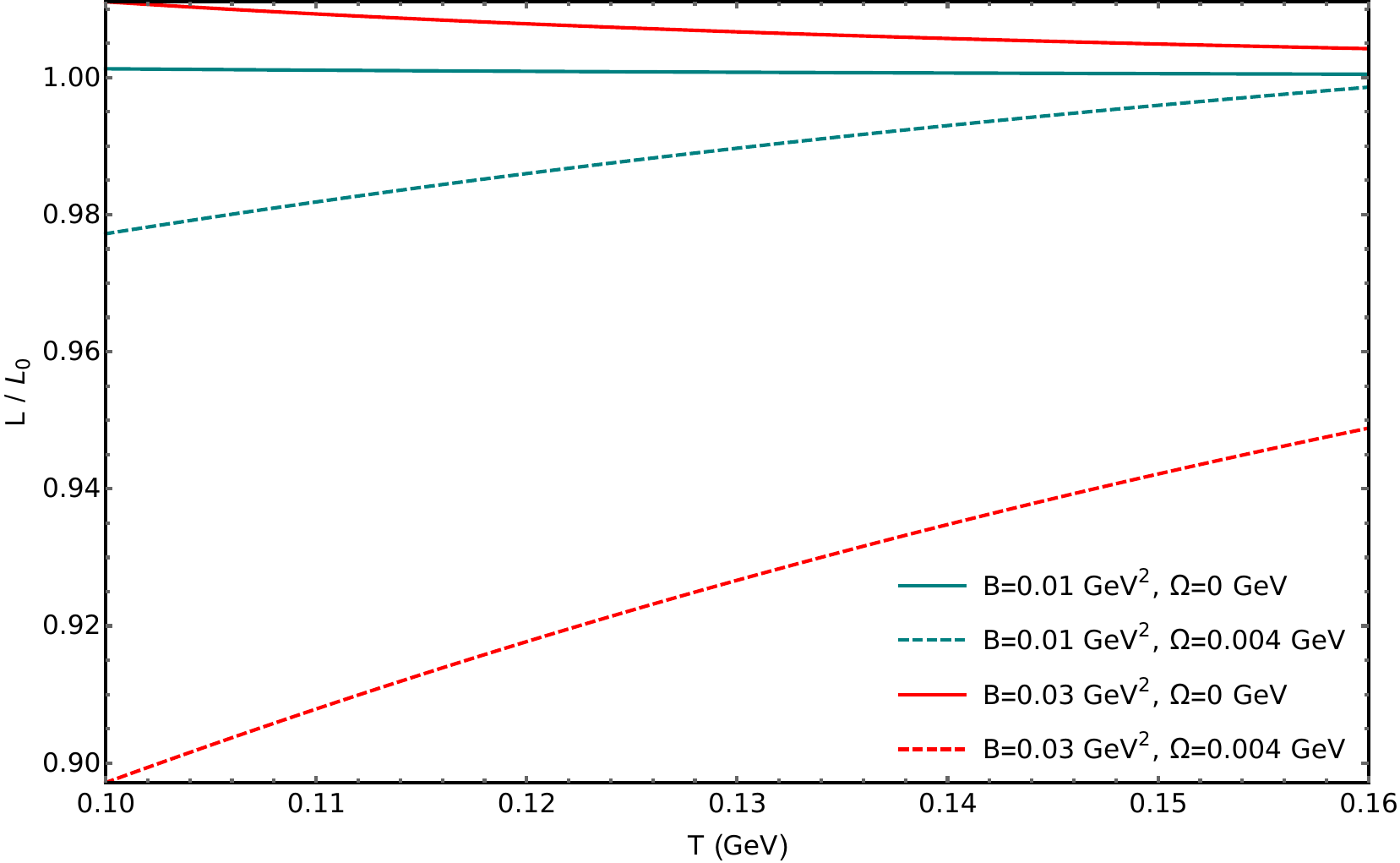}
        \caption{}
        \label{NormL/T}
    \end{subfigure}
    \begin{subfigure}[b]{0.45\textwidth}
        \centering
        \includegraphics[width=\textwidth]{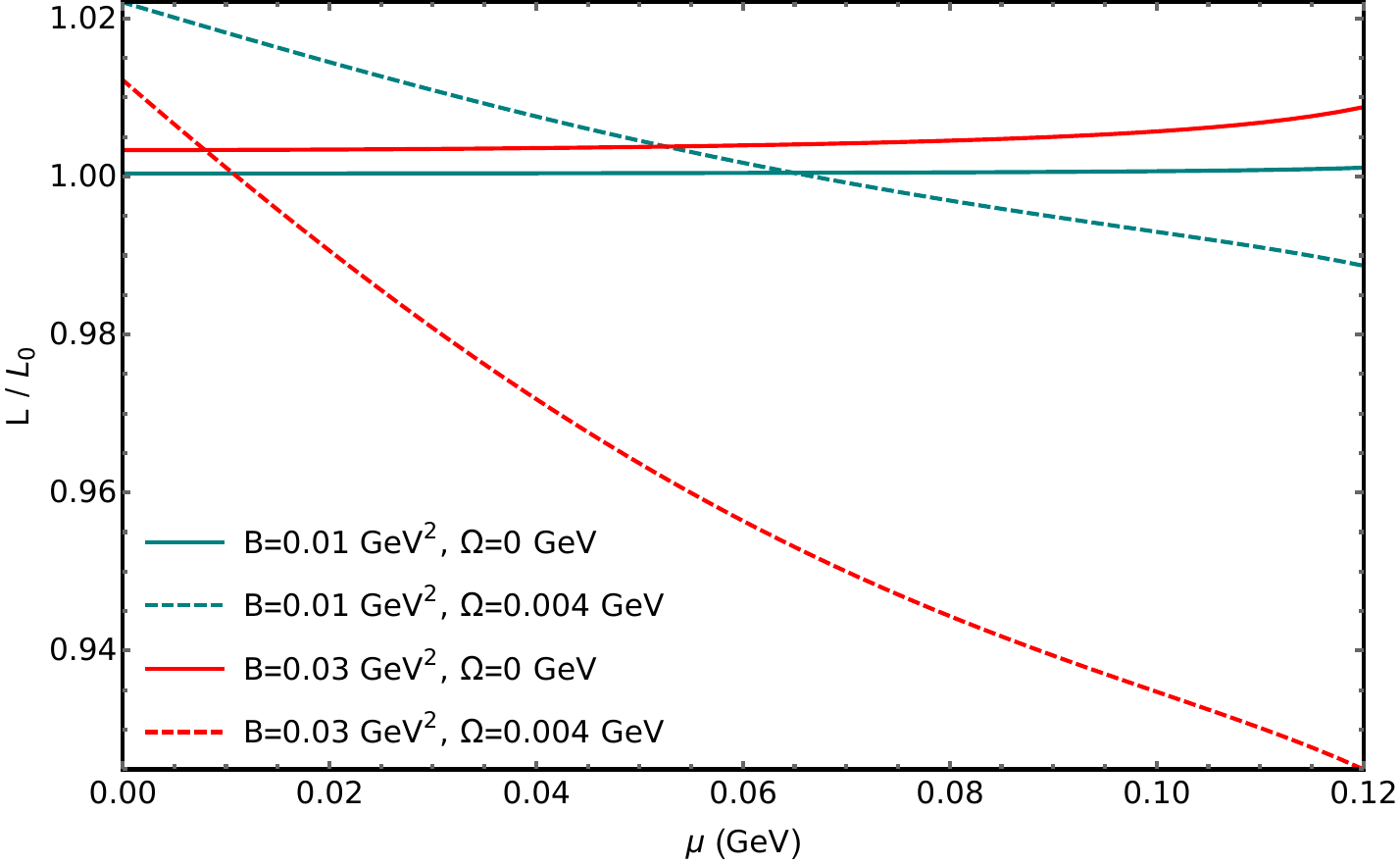}
        \caption{}
        \label{NormL/mu}    
    \end{subfigure}
    \caption{\justifying \small Normalized Lorenz number for $R=50\,\,\text{GeV}^{-1}$ with (a) magnetic field ($T=0.14\,\,\text{GeV}$ and $\mu=0.1\,\,\text{GeV}$), (b) angular velocity of rotation ($T=0.14\,\,\text{GeV}$ and $\mu=0.1\,\,\text{GeV}$), (c) temperature $T$ ($\mu=0.1\,\,\text{GeV}$), and (d) charge chemical potential $\mu$ ($T=0.14\,\,\text{GeV}$). Increasing the magnetic field strictly suppresses $L/L_0$, while angular velocity induces a non-monotonic response. 
    Finite rotation qualitatively changes the temperature and chemical-potential dependence of the Lorenz ratio, for a static ($\Omega=0$) medium: 
    Instead of $L/L_0$ decreasing with temperature and increasing with chemical potential, rotation causes it to increase with temperature and decrease with chemical potential.}
    \label{NormL/BomegaTmu}
\end{figure*}

We have shown the variation of the longitudinal thermal conductivity with magnetic field and angular velocity of rotation in Figs.~\ref{kappa/B} and \ref{kappa/omega}, respectively. The behavior of thermal conductivity with $B$ and $\Omega$ is almost identical to that of electrical conductivity, which is expected as both coefficients are determined by the same longitudinal phase-space integral weighted by $(p_z/p_{0j})^2$ and $f_0(1+f_0)$, with only the energy vs enthalpy weighting distinguishing them. The neutral pion contribution to the thermal conductivity $\kappa_{\parallel}^{\text{neutral}}$, being independent of $B$, is responsible for the overall offset of the curves. Figs.~\ref{S/B} and \ref{S/omega} show the variation of the Seebeck coefficient $S$ with $B$ and $\Omega$, respectively. Again, the behavior resembles that of electrical and thermal conductivities for the same reason. The negative sign is the consequence of our convention ($S=\frac{E_z}{dT/dz}$ with conserved charge $q>0$ for $\pi^+$): for the chemical potential $\mu=0.1\,\,\text{GeV}$ used in Fig.~\ref{TC/Bomega}, the $\pi^+$ population dominates and the diffusive flow of the more numerous $\pi^+$ down the temperature gradient is balanced by an electric field oriented opposite to the gradient.

Fig.~\ref{sigma/mu} shows the variation of $\sigma_{\parallel}/T$ with the charge chemical potential $\mu$ for different values of $B$ and $\Omega$. We see an increasing trend of electrical conductivity with $\mu$. An interesting point to note is that when $\Omega=0$, the magnetic-field effect on electrical conductivity changes little with increasing charge chemical potential; however, when we have a non-zero $\Omega$, the magnetic field effect is more pronounced at higher values of $\mu$, as compared to lower values of $\mu$. A similar trend is seen for the case of thermal conductivity in Fig.~\ref{kappa/mu}. However, rotation seems to have a stronger effect on thermal conductivity than the magnetic field does, compared to the case of electrical conductivity. In Fig.~\ref{S/mu}, Seebeck coefficient $S$ has been plotted against $\mu$ for different values of $B$ and $\Omega$. Here also, we see an increasing trend in the magnitude of $S$ with $\mu$. When $\Omega=0$, $S$ vanishes at zero chemical potential, which is what we expect since at $\mu=0$, the system is perfectly charge symmetric and thermal fluctuations in $\pi^+$ and $\pi^-$ will exactly cancel out; therefore, no charge current is generated by the temperature gradient. However, for non-zero $\Omega$, even at $\mu=0$, we see a non zero $S$. This indicates that $j\Omega l$ in the spectrum of charged pions plays the role of a chemical potential. This has also been discussed in Ref.~\cite{zahed}.

Figs.~\ref{sigma/T}, \ref{kappa/T}, and \ref{S/T} show the variation of the coefficients with temperature for different values of $B$ and $\Omega$. The decreasing trend of $\sigma_{\parallel}/T$ with $T$ is driven primarily by the explicit $1/T^3$ scaling of the relaxation time, which more than compensates the thermal increase in the number of available carriers. The qualitative behavior is consistent with Ref.~\cite{PhysRevC.107.034903}. By contrast, $T\kappa_{\parallel}$ rises with $T$. The qualitative behavior and the range of values of $T\kappa_{\parallel}$ with $T$ are consistent with Ref.~\cite{PhysRevD.89.054013}. The Seebeck coefficient remains relatively flat with temperature in this window, indicating that the proportional generation of thermoelectric current maintains a stable balance between the thermal and electrical gradients, irrespective of the absolute temperature. In all these plots, we see a significant enhancement in the coefficients due to rotation across the temperature window.

The variation of the normalized Lorenz ratio $L/L_0$ with $B$, $\Omega$, $T$, and $\mu$ is shown in Fig.~\ref{NormL/BomegaTmu}. It is important to note that the electrical conductivity $\sigma_{\parallel}$ is purely mediated by charged pions, whereas thermal conductivity $\kappa_{\parallel}$ receives contributions from both charged and neutral pions. As shown in Fig.~\ref{NormL/B}, the introduction of finite angular velocity $\Omega$ causes $L/L_0$ to strictly decrease with the increase in the magnetic field. This suppression indicates that the combination of rotation and magnetic field enhances charge flow more rapidly than heat flow. The behavior of $L/L_0$ is non-monotonic with angular velocity $\Omega$, as depicted in Fig.~\ref{NormL/omega}. At low values of $\Omega$, the ratio initially dips, indicating the overpowering of charge transport over heat transport. However, at higher values of $\Omega$, $L/L_0$ starts to increase again. This behavior arises from the fundamentally different phase-space weighting of charge and heat transport. Fig.~\ref{NormL/T} shows an increase in the ratio $L/L_0$ with temperature when there is non-zero rotation, whereas in the absence of rotation, we see a slight decrease in $L/L_0$ with $T$. This shows that, in the absence of rotation, an increase in $T$ favors charge transport more than the heat transport; however, the opposite happens when rotation is switched on. The effect of the charge chemical potential $\mu$ on $L/L_0$ contrasts that of $T$ on $L/L_0$, as can be seen in Fig.~\ref{NormL/mu}. In the absence of rotation, with the increase in $\mu$, the heat transport slightly overpowers charge transport, whereas when $\Omega\ne0$, charge transport dominates over heat transport. To obtain the plots in Fig. \ref{sigma/B}, \ref{sigma/omega}, \ref{sigma/mu}, and \ref{sigma/T}, we have considered $(l_{\text{max}},k_{\text{max}})=(180,180)$. For all other plots, we have taken $(l_{\text{max}},k_{\text{max}})=(160,160)$, $(l^u_{\text{max}},k^u_{\text{max}})=(170,170)$.

\section{Conclusion and Outlook}\label{outlook}
In this work, we investigated the electric, thermal, and thermoelectric responses of a rotating pion gas of finite radius in the presence of a constant background magnetic field, with the axis of rotation aligned with the magnetic field. We discussed pion condensation due to rotation, and demonstrated that within the range of magnetic fields, angular velocities, and finite-size parameters explored in this work, no $\pi^-$ condensate was found. For $\pi^+$, we explicitly determined the condensation thresholds within this space and restricted our operational parameters to remain well outside them, ensuring our transport analysis is strictly free of condensation effects.
Using the Boltzmann transport equation in the relaxation-time approximation, we derived the longitudinal electrical conductivity, thermal conductivity, Seebeck coefficient, and Lorenz number. Since the magnetic field quantizes the transverse motion of charged pions, the transport analysis was restricted to the direction parallel to the magnetic field and the rotation axis. Our results show that the magnetic field and rotation have competing effects on longitudinal transport. In the absence of rotation, the magnetic field suppresses the electrical conductivity by increasing the effective energy of charged pion modes and reducing the available thermal population of charge carriers. Rotation, however, lowers the energy of angular-momentum-carrying modes through the rotational shift in the spectrum. As a result, sufficiently strong rotation can overcome the magnetic suppression and reverse the magnetic-field dependence of the electrical conductivity. Similar qualitative behavior is found for the thermal conductivity and the Seebeck coefficient.

The thermoelectric response displays a notable feature. At zero rotation and vanishing charge chemical potential, the Seebeck coefficient vanishes because of the exact symmetry between positive and negative pions. In contrast, when rotation is present, a finite Seebeck coefficient can arise even at zero charge chemical potential. This indicates that the coupling between rotation and angular momentum acts as an effective chemical potential.

We also studied the normalized Lorenz ratio, $L/L_0$ to understand how the magnetic field and rotation affect the relative strengths of heat and charge transport. We found that the relative significance of heat and charge transport is sensitive to both magnetic field and rotation velocity. In a rotating medium, increasing the magnetic field strictly suppresses the relative significance of heat transport. This transport balance also responds non-monotonically to angular velocity, with charge transport initially outpacing heat transport at low rotation rates before heat transport starts increasing at higher rotation rates. Notably, finite rotation fundamentally reverses the system's thermal and chemical dependencies: while the relative significance of heat-to-charge transport typically decreases with temperature and increases with chemical potential, rotation completely inverts both of these trends.

In summary, our study demonstrates that finite-size effects, rotation, and magnetic field induced Landau quantization collectively generate a crucial interplay in the transport properties of a pion gas, leading to qualitative modifications of electrical, thermal, and thermoelectric responses that may be relevant for strongly magnetized and vortical matter produced in heavy-ion collisions.

The framework established in this study lays the groundwork for several critical extensions in high-energy physics. Most notably, a natural progression is to extend this formalism beyond hadronic matter to a complete quark-gluon system, enabling the exploration of hydrodynamic evolution within the QGP. To achieve a comprehensive picture of dissipative dynamics in these extreme environments, future work can move beyond the RTA by implementing more realistic collision kernels and accounting for how magnetic fields and rotation modify scattering cross-sections and the associated relaxation time ($\tau_R$). Furthermore, this framework may be applied to study heavy quark transport across both hadronic and QGP media. Finally, bridging these theoretical advancements with phenomenology and experimental data is paramount; future research may connect the modified conductivities to measurable heavy-ion collision observables, such as dilepton emission rates and the elliptic flow coefficient ($v_2$).

\section{Acknowledgment}
We thank Sourav Dey for a helpful discussion. D.G. (JRF Code: 231610146577) is supported by the Junior Research Fellowship of the
University Grants Commission, India. VC acknowledges the Anusandhan National Research Foundation (ANRF) for Advanced Research Grant (ARG) ( Grant No.:  ANRF/ARG/2025/002424/PS).

\bibliography{ref}{}

\end{document}